\documentclass[apj]{emulateapj}
\usepackage{xspace}
\usepackage{graphicx}
\usepackage{natbib}
\usepackage{subfigure}
\usepackage{longtable}
\usepackage{nicefrac} 
\usepackage{amsmath}
\usepackage{hyperref}
\usepackage{threeparttable}
\usepackage{tablefootnote}
\usepackage{threeparttablex}
\usepackage{afterpage} 
\usepackage{lineno}
\usepackage{booktabs}
\newcommand{\ergcm}{\ensuremath{\,{\rm erg\,cm^{-2}}}\xspace}

\newcommand{\dm}{\ensuremath{\,{\rm pc\,cm^{-3}}}\xspace}

\newcommand{\rat}{\ensuremath{\,{\rm Jy  \ ms \ erg^{-1} \ cm^2}}\xspace}
\begin{document}

\title{A Search for High-Energy Counterparts to Fast Radio Bursts}

\author{Virginia Cunningham\altaffilmark{1}, S. Bradley Cenko\altaffilmark{2,3}, Eric Burns\altaffilmark{2}, Adam Goldstein\altaffilmark{4}, Amy Lien\altaffilmark{2}, Daniel Kocevski\altaffilmark{5}, Michael Briggs\altaffilmark{6}, Valerie Connaughton\altaffilmark{4}, M. Coleman Miller\altaffilmark{1,3}, Judith Racusin\altaffilmark{2}, and Matthew Stanbro\altaffilmark{6}}

\altaffiltext{1}{Astronomy Department, University of Maryland, College Park, MD 20742, USA}
\altaffiltext{2}{Astrophysics Science Division, NASA/Goddard Space Flight Center, MC 661, Greenbelt, MD 20771, USA}
\altaffiltext{3}{Joint Space-Science Institute, University of Maryland, College Park, MD 20742-2421, USA}
\altaffiltext{4}{Science and Technology Institute, Universities Space Research Association, Huntsville, AL, USA}
\altaffiltext{5}{Astrophysics Office, ST12, NASA/Marshall Space Flight Center, Huntsville, AL 35812, USA}
\altaffiltext{6}{Department of Space Science, University of Alabama in Huntsville, Huntsville, AL, USA}

\begin{abstract}
We report on a search for high-energy counterparts to fast radio bursts (FRBs) with the \textit{Fermi} Gamma-ray Burst Monitor (GBM), \textit{Fermi} Large Area Telescope (LAT), and the \textit{Neil Gehrels Swift Observatory} Burst Alert Telescope (BAT). We find no significant associations for any of the 23 FRBs in our sample, but report upper limits to the high-energy fluence for each on timescales of 0.1, 1, 10, and 100 s. We report lower limits on the ratio of the radio to high-energy fluence, $\nicefrac{f_{r}}{f_{\gamma}}$, for timescales of 0.1 and 100 s. We discuss the implications of our non-detections on various proposed progenitor models for FRBs, including analogs of giant pulses from the Crab pulsar and hyperflares from magnetars. This work demonstrates the utility of analyses of high-energy data for FRBs in tracking down the nature of these elusive sources. \\
\end{abstract}


\section{Introduction\label{sec:intro}}

\indent Fast radio bursts (FRBs) are bright (typical fluences of 2 Jy ms), short-duration (few ms) pulses at frequencies of $\sim$1 GHz \citep{Lorimer2007, Thornton2013}. FRBs can be distinguished from other short-duration radio pulses (e.g., pulsars) by their high dispersion measures (DM) for their Galactic latitude (100 to 2600$\dm$; \citealt{Petroff2016}). Because the DM derived for FRBs can be significantly in excess of the Galactic value (average of $\sim$ $250 \dm$ for Galactic pulsars; \citealt{Manchester2016}), they must either reside in regions of large over-densities of free electrons if in the Milky Way or at extragalactic distances.\\
\indent The first FRB was discovered in 2007 \citep{Lorimer2007} and it was not until 2013 that their reality as a class of astrophysical objects was firmly established (\citealt{Thornton2013}, c.f. perytons). Only $\sim$70 FRBs have been published in the literature at the date of this publication (see the FRB Catalog (FRBCAT) at \url{frbcat.org}; \citealt{Petroff2016}). However, because these have been discovered by relatively narrow field-of-view instruments, the true all-sky rate is remarkable: \mbox{$\sim6000 \ \rm sky^{-1} \ \rm day^{-1}$} above a fluence of $\sim$few Jy ms at $\sim$1 GHz \citep{KeanePetroff2015, Champion2016, Nicholl2017}. For comparison, this is much larger than the all-sky rate of gamma-ray bursts ($\sim$few per day at current detector sensitivities) and comparable to the rate of supernovae (core-collapse and thermonuclear) out to a redshift of $z\approx$ 0.5 \citep{Li2011}.\\
\indent Only two FRBs are known to exist as repeating bursts: FRB121102 (``The Repeater") and FRB180814.J0422+73 \citep{Spitler2016,Amiri2019}. FRB180814.J0422+73 was reported as this work was being completed and so is not included in our following analysis. \cite{Spitler2016} report the detection of 12 bursts from the Repeater at 1.4 GHz from Arecibo and 5 bursts at 2 GHz from Green Bank, \cite{Chatterjee2017} report 9 bursts at 3 GHz from the Very Large Array (VLA), and \cite{Scholz2017} report 8 bursts at 2 GHz from Green Bank, 2 bursts at 1380 MHz from Arecibo, and 2 bursts seen by both telescopes. All repeating bursts display a consistent DM but can vary in pulse shape and spectral shape. The Repeater exhibits no evidence for periodicity, but instead appears to show episodes of enhanced activity (i.e., active and quiescent periods). Other FRBs have been reobserved, but none show repeated bursting as displayed by the two previously mentioned. In several cases it is possible to rule out repeat outbursts with the intensity and frequency of FRB121102; however less frequent and/or fainter, repeated outbursts remain viable \citep{Palaniswamy2018}. It is therefore not currently known if all FRBs repeat, or if the known population comprises multiple classes of events (e.g., repeaters and non-repeaters).\\
\indent Only the repeating FRB121102 has been localized to within a host galaxy. No obvious host has yet been identified for FRB180814.J0422+73. \cite{Chatterjee2017} use high angular resolution radio interferometry to place FRB121102 within a dwarf galaxy at $z\sim$ 0.2 \citep{Tendulkar2017}. The FRB location is consistent with a faint, persistent radio source of unknown origin \citep{Chatterjee2017}. While this result provides unprecedented insight into the physics of the repeating FRB, without detections from radio interferometers for the other FRBs it is impossible to localize them to such high precision using this method. \\
\indent Despite being a recent discovery, FRBs have nonetheless piqued great interest in the area of high time-resolution radio astronomy. This excitement can be divided into two separate motivations: FRBs may become powerful probes of the intergalactic medium (IGM), and the emission mechanism powering these outbursts may help clarify some long-standing issues in astrophysics, including the missing baryon problem and the nature of coherent emission (see below).\\
\indent The large DM derived from FRBs suggest that the signals have encountered more free electrons than can be accounted for in the ISM of the Milky Way. While Galactic models resulting from a large local density of free electrons do exist (e.g., \citealt{Maoz2015}), the most natural explanation is that FRBs are of extragalactic origin (e.g., \citealt{Thornton2013}). In fact if all the DM were to result from electrons in the intergalactic medium (IGM), this would imply cosmological redshifts of $z\approx$0.5-1.0 for these events.\\
\indent The possibility of using FRBs to measure the density of free electrons at cosmological distances may offer a way to solve the ``missing baryons" problem. In the local universe, only half of cosmic baryons reside at densities and temperatures that result in detectable emission and/or absorption \citep{Bregman2007, Shull2012}. \cite{McQuinn2014} demonstrates how the location of these baryons can be inferred from the distribution of DM (at a fixed redshift). Similarly, samples extending out to $z\approx$ 3 with measured DM and redshift may even be able to constrain the equation of state of dark energy \citep{Zhou2014}.\\
\indent In addition to their potential utility as cosmological probes, FRBs also offer a means to improve our understanding of coherent emission processes. Any source emitting incoherently (e.g., synchrotron radiation) cannot exceed a brightness temperature of $10^{12}$ K \citep{Readhead1994}. For a typical FRB with an intrinsic duration of 1 ms, causality limits the size of the emitting region to be less than 300 km barring bulk relativistic motion. For FRBs at distances of $\sim$1 Gpc, the peak flux densities of $\sim$1 Jy at $\nu \sim$1 GHz imply a brightness temperature of $\rm T_B \gtrsim 10^{35}$ K. Clearly for FRBs $\rm T_B \gg 10^{12}$ K, from which we infer that FRBs must be emitting coherent radiation. Only a handful of astronomical sources are known to radiate coherently, with pulsars being the most well-known example. Given the large uncertainties in the pulsar emission mechanism, the advent of FRBs offer the real hope of fundamental progress towards understanding coherent processes in this long-standing field.\\
\indent As with many astronomical phenomena, the number of theoretical models has rapidly grown larger than the number of known FRBs. Here we consider several of the more plausible models, which must incorporate the following basic tenets: compact emission region, extragalactic distance scale, coherent emission mechanism, repeated outbursts from at least some FRBs, and large all-sky rates. We consider models for FRBs resulting from outbursting neutron stars (either magnetically or rotationally powered), as mergers between neutron star binaries, or as ``cosmic combs". We describe the models and their various predictions in greater detail in $\S$\ref{sec:analysis}.\\
\indent The goal of this paper is to search for possible counterparts at high-energy wavelengths to FRBs. We use data from the \textit{Fermi} Gamma-ray Burst Monitor \citep[GBM;][]{Meegan2009}, the \textit{Fermi} Large Area Telescope \citep[LAT;][]{Atwood2009}, and the \textit{Neil Gehrels Swift Observatory} \citep{Gehrels2004} Burst Alert Telescope \citep[BAT;][]{Barthelmy2005} to search for X-ray and gamma-ray (8 keV to 300 GeV) counterparts to FRBs. Although the energy range of the \textit{Swift} BAT overlaps with the \textit{Fermi} GBM we choose to include the BAT due to its arcminute localization, compared to the GBM. \cite{Scholz2016} use the same instruments to search for sources related to the repeating FRB but report no significant detections. They conducted another campaign coordinating observations between the Green Bank, Effelsberg, and Arecibo radio telescopes and the \textit{Chandra X-ray Observatory} and \textit{XMM-Newton} \citep{Scholz2017b} but also report no significant X-ray detections. Their searches focus on a single FRB but our project extends to cover all FRBs within the field-of-view of each instrument as well as extending the timescales of interest that were analyzed. One advantage of this population study is the ability to potentially identify fundamental differences between repeating and non-repeating FRBs. There also exist upper limits for three FRBs from the INTEGRAL observatory \citep{2018ATel11431,2018ATel11387,2018ATel11386} which has comparable energy coverage to the \textit{Fermi} GBM and \textit{Swift} BAT. These limits are in agreement with the limits found here in this paper.\\
\indent \cite{Tendulkar2016} place limits on the ratio of radio to gamma-ray emission for FRBs based on observations of SGR1806$-$20. We conduct a more sensitive search for high-energy counterparts in the GBM by employing the targeted search techniques developed for coincident searches for gravitational wave counterparts \citep{blackburn2015high}. We also use these ratios to compare our results with the proposed gamma-ray counterpart to FRB131104 \citep{DeLaunay2016}. With the exception of the host galaxy for FRB 121102 \citep{Chatterjee2017, Tendulkar2017, Marcote2017}, no other electromagnetic analogs have so far been confirmed, despite rigorous efforts. \\
\indent The detection of robust high-energy signals from FRBs would have a significant impact on the field as current theories predict widely differing high-energy fluences. Although a confirmed, positive detection of a high-energy counterpart would definitively rule out many theories, a non-detection and corresponding upper limit could also eliminate many as well. \\
\indent This paper is organized as follows: in $\S \ref{sec:data}$ we describe the data products and analysis methods for calculating the high-energy upper limits for each FRB. In $\S \ref{sec:analysis}$ we compare our results with various theories from the literature before we draw our conclusions in $\S \ref{sec:conclusions}$ on the likelihood and implications for each model. In this work we assume a standard $\Lambda$CDM cosmology and that the Milky Way is well-described by Galactic structure models such as NE2001 \citep{CordesLazio2002}. \\

\section{Data and Results} \label{sec:data}

\indent There are 23 published FRBs used in this analysis (taken from the FRBCAT) as of July 2017. 17 were detected with the Parkes Radio Telescope, 3 with UTMOST (Upgrade of The Molonglo Observatory Synthesis Telescope), 1 with the Arecibo Telescope, 1 with the Green Bank Telescope, and 1 with ASKAP (the Australian Square Kilometre Array Pathfinder). We search for contemporaneous high-energy emission from all these events with three different instruments: \textit{Fermi} GBM, \textit{Fermi} LAT, and \textit{Swift} BAT (see Table \ref{tab:FRBs} in the appendix for a breakdown of available observations per FRB). We search for high-energy emission on a variety of different timescales. To place limits on a coincident (i.e., ms-long) pulse, we utilize the smallest time bin available from each relevant instrument. Where possible, we also place limits on timescales\footnote{We take the zero-point time, $t_{frb}$, as the arrival time of an infinite energy photon.} of \mbox{0-1 s}, 0-10 s, and 0-100 s. This spans the range from hyperflares of magnetars ($\Delta t \sim 0.1$ s) to short (\mbox{$\Delta t \sim 1$ s}) and long ($\Delta t \sim 10-100$ s) gamma-ray bursts (GRBs).  \\

\subsection{\textit{Fermi} GBM \label{sec:GBM}}

The Gamma-ray Burst Monitor \citep[GBM;][]{Meegan2009} is a collection of hard-X-ray/soft gamma-ray detectors onboard the \textit{Fermi Gamma-ray Space Telescope} sensitive to photons with energies from 8 keV to 40 MeV. \textit{Fermi} is in a low-Earth (96 min) orbit, and the GBM is sensitive to gamma rays from the entire sky unocculted by Earth when outside the South Atlantic Anomaly (SAA).\\
\indent The GBM consists of two sets of detectors: twelve sodium iodide (NaI) scintillators cover a lower energy range from 8 keV to 1 MeV, and two bismuth germanate (BGO) scintillators cover the higher end from 300 keV to 40 MeV. The 12 NaI detectors are positioned to enable all-sky coverage, while the two omnidirectional BGO detectors are positioned on opposite sides of the spacecraft for the same reason. The 14 detectors are positioned in such a way that any burst should be seen by multiple detectors. The 12 NaI detectors are used for triggering and localization and the two BGO enable a broader energy range for spectroscopy. The rates received by each detector combined with their relative position and angle to each other allow the position of bursts to be determined to a few degrees accuracy \citep{Connaughton2015}. \\
\indent Each of the 14 detectors in the GBM record several data products. The two of interest for this work are continuous time (CTIME) and time-tagged events (TTE). The CTIME data are binned by 0.256 seconds with eight energy channels. The TTE data are continuous event data precise to 2 microseconds with 128 energy channels. Due to the short duration of FRBs, TTE data are preferred over CTIME; however continuous TTE data only started in 2012, so are not available for every FRB in our sample.\\
\indent Of the 23 FRBs in our sample, 20 occurred after \textit{Fermi}'s launch. Of those 20, 12 were visible to \textit{Fermi} during good time intervals for GBM. Of the 38 repeat bursts of FRB 121102, 15 were visible to \textit{Fermi} during good time intervals for GBM. To determine if a candidate counterpart exists in GBM data we ran a targeted search \citep{blackburn2015high, 2016arXiv161202395G} of GBM data around $t_{frb}$ for $\pm$15 s for the 0.1 and \mbox{1.0 s} timescales, $\pm$250 s for the 10 s timescales, and $\pm$400 s for the 100 s timescales (the 100 s timescale was only searched when the background was stable over periods of a few hundred seconds and had continuous TTE coverage). \\
\indent The targeted search was designed to identify untriggered, faint, short GRB-like counterparts to gravitational-wave events, which makes it a useful tool to adapt to our purposes. We use the same three standard spectral templates described in \citet{2016arXiv161202395G}, which generally cover the diverse range of short to long GRBs: a low-energy soft Band function (\citealt{Band1993}; $E_{\rm peak}=70$ keV), a medium-energy Band function ($E_{\rm peak}=230$ keV), and a power law with an exponential cutoff ($E_{\rm peak}=1.5$ MeV). While we calculate fluence upper limits for each of these three spectral types, the limits listed in this paper will be given for the hardest template. On average, this harder spectral template results in a factor of $\sim$2.5 times the fluence of the medium-energy template and $\sim$5 times the fluence of the low-energy template. \\
\indent We employ the Bayesian likelihood analysis originally developed by \cite{blackburn2015high} to search for contemporaneous signals around the FRB radio detections. This method calculates the likelihood of a signal matching one of the three spectral templates compared to the null hypothesis of a constant background. Owing to the highly transient universe in the gamma-ray band and GBM's all-sky coverage there were a few real transient gamma-ray signals in GBM during time intervals of interest; however, these are known to be unrelated due to inconsistent sky localizations or classification as a known source type (e.g., a solar flare). No possibly related signal is significant over the total lifetime of the search (See the Appendix for more detail on these unrelated signals). \\
\indent In the absence of any correlated gamma-ray signal with the FRBs, we estimate flux upper limits in the search time windows around each $t_{frb}$ on timescales of 0.1, 1, 10, and 100 s using the same spectral templates that were used by the targeted search.  These conservative upper limits were calculated by utilizing the NaI detector with the smallest normal angle to the FRB, and estimating the maximum $3 \sigma$ count rate flux upper limit in the window based on the modeled background noise.  The count rate upper limit was then converted to a flux upper limit by assuming each of the template spectra, folding them through the GBM detector responses calculated for the FRB sky location, and fitting for the amplitude of the template spectra.  This procedure results in $3\sigma$ flux upper limits listed in Table~\ref{tablat}. Five of these FRBs are analyzed by \cite{Tendulkar2016} where the limiting gamma-ray fluence is estimated to be $1\times 10^{-8} \ergcm$, roughly consistent with the faintest known short GRBs detected by GBM. The targeted search used here provides consistent, though slightly shallower, limits to \cite{Tendulkar2016}. \\
\indent In addition, we consider the results derived from performing a stacking analysis of the bursts from the Repeater and a separate stacking analysis of the bursts from the non-repeating FRBs. In the case of the non-repeating FRBs we assume that all FRBs are approximately at the same redshift (\mbox{z $=0.1-0.3$}). This assumption will be justified in $\S$\ref{sec:analysis} where each of the models we consider in this work limits the distance of the FRBs to no further than $\sim$1 Gpc. We find no obvious potential signals which would warrant any further stacking analysis for either case. \\
\subsection{\textit{Fermi} LAT}
\indent The Large Area Telescope \citep[LAT;][]{Atwood2009} is a pair-conversion telescope on board the \textit{Fermi} satellite, sensitive to gamma rays with energies between 20 MeV and more than 300 GeV. The LAT has a wide field-of-view (FOV), scans continuously and covers approximately 20\% of the sky at any given time. The LAT completes all-sky coverage every two orbits over a duration of about three hours.  The timing accuracy of the LAT is better than 10 $\mu$s and its localization precision is highly energy-dependent ($\sim$ $5^{\prime}$ for GeV photons). \\
\indent We search the \textit{Fermi} LAT data for gamma-ray counterparts by performing an unbinned likelihood analysis using the standard analysis tools developed by the LAT team (ScienceTools version v10r01p0)\footnote{http://fermi.gsfc.nasa.gov/ssc/}. For this analysis, we use the `P8R2\_TRANSIENT\_V6' instrument response functions and select `Transient' class events in the 0.1--100 GeV energy range from a $12^\circ$ radius energy-independent region of interest (ROI) centered on the FRB location. The size of the ROI is chosen to reflect the 95\% containment radius of the LAT energy-dependent point spread function at 100 MeV.  The `Transient' event class is chosen because it represents looser cuts against non-photon background contamination and is typically used to study GRBs on very short timescales \citep{LATPerformancePaper}. \\
\indent In standard unbinned likelihood analysis, the observed distribution of counts at a particular position is fit to a model that includes all known gamma-ray sources in the 3FGL catalog \citep{Acero2015} within a radius of 30$^\circ$, as well as Galactic and isotropic background components \footnote{https://fermi.gsfc.nasa.gov/ssc/data/access/lat/BackgroundModels.html}. The Galactic component, \emph{gll\_iem\_v06}, is a spatial and spectral template that accounts for interstellar diffuse gamma-ray emission from the Milky Way.  The isotropic component, \emph{iso\_transient\_v06}, provides a spectral template to account for all remaining isotropic emission including contributions from both residual charged particle backgrounds and the isotropic celestial gamma-ray emission. Possible emission from a FRB is modeled as an uncatalogued point source with a power law spectrum where the normalization and photon index are left as free parameters. A likelihood-ratio test is then employed to quantify whether there exists a significant excess of counts due to the uncatalogued point source above the expected background model. If no significant new source is found, we calculate the 95$\%$ confidence level upper limits using a Bayesian method described in \citet{Ackermann2016}, which we convert to a fluence limit for the relevant timescale. Note that these fluence limits are calculated via a different method than we use for the GBM ($\S$2.1) and BAT ($\S$2.3) data.\\
\indent The three earliest FRBs are again excluded from our analysis since they occurred before \textit{Fermi} was launched on June 11, 2008. Of the remaining 19 non-repeating FRBs, six are located within the LAT FOV at the time of radio detection. Five of the 38 repeating bursts are in the LAT FOV as well. We examine two time intervals based on the zero-point detection time, $t_{FRB}$: $0-10$ s and $0-100$ s (Table \ref{tablat}). No photons are detected for any of the FRBs within 1 second of the initial burst. \\

\subsection{\textit{Swift} BAT} \label{sec:BAT}
\indent The \textit{Neil Gehrels Swift Observatory} \citep{Gehrels2004} Burst Alert Telescope (BAT) is a coded-aperture instrument dedicated to triggered hard X-ray observations of GRBs. The BAT detectors have an energy range of 15 to 300 keV with a resolution of $\sim$7 keV, a large FOV of 1.4 steradians (half-coded) and a positional accuracy of $\sim$3$^{\prime}$ \citep{Barthelmy2005}. Although the detectors are sensitive up to 300 keV, the coded mask is transparent to photons above 150 keV and so is unable to determine their direction from the sky. When running in survey mode, BAT collects detector plane histograms that are binned in $\sim$300 s. These detector plane histograms can be used to generate sky images and search for sources in the BAT FOV. In addition to these spatially-resolved images, BAT also collects raw rate data from all of the enabled detectors. The raw rate data are a continuous stream of events which can be used to search for GRB triggers not in the BAT FOV. We analyze both the five-minute time-binned survey images and the 64 ms-binned, four energy band (15-25 keV, 25-50 keV, 50-100 keV, and 100-350 keV) rate data lightcurves using the standard \textit{Swift}-specific tools provided by the HEASoft package (version 6.18). \\
\indent Only FRB110626, FRB150215, and FRB160410 were within the BAT FOV at the time of radio detection. The three earliest FRBs occured prior to \textit{Swift}'s launch on November 20, 2004 so they are excluded from the analysis. Of the 19 non-repeating FRBs examined 10 of the bursts were out of the FOV, and one did not occur during recorded observations (i.e., the telescope was most likely slewing to a new location), 3 were within the FOV while the BAT was in the SAA, and 1 occurred while the BAT was slewing. 30 of the 38 repeating signals from FRB121102 were not within the FOV at the time of detection and six occured while the BAT was in the SAA. Three of the bursts (FRB131104 and bursts 2 and 3 of FRB121102) were located right on the edge of the BAT FOV but are excluded from analysis due to their low partial coding fraction. It is standard practice\footnote{\cite{DeLaunay2016} report a \textit{Swift} BAT counterpart to FRB131104 with a partial coding fraction of 2.9\%. A more detailed analysis of this event is currently underway for a separate work (Sakamoto et al., in prep).} to remove pointings with partial coding fractions corresponding to less than 10\% of the array being illuminated \citep{Krimm2013}.\\
\indent The survey images provide more accurate positional information compared to the rate data. The rate data are the cumulative sum of all counts seen within and around the BAT FOV. It can be difficult to definitively attribute a significant rise in counts to any single FRB as it could also be due to a nearby source. In the survey images the precise timing information is lost due to the 300-second binning of the counts. While the rate data are useful for identifying sudden significant changes in the aggregate background emission on short timescales, the survey images are more accurate for producing limits for the specific FRB locations. The fluence limits derived from the rate data are shallower than, yet still consistent with, the limits derived from the survey images. \\
\indent We find no significant counterpart detections at 3$\sigma$ confidence level for FRB110626, FRB150215, and FRB160410, but we are able to determine upper limits to the high-energy fluence. We produce 8-channel spectra using the mask-weighted background variation counts detected in the survey images, and estimate the flux that would have been equivalent to a 3$\sigma$ detection. Assuming a simple power law function with index 2.0 for the FRB spectra, we calculate an estimate of the flux within XSPEC based on the spectral fit over an energy range of 15 to 350 keV \citep{Arnaud1996}. We use the FRB location on the BAT FOV to generate the instrument response matrix corresponding to the respective grid ID on the BAT detector \citep{Lien2014}. The fluence limits are listed in Table \ref{tablat} for 300-second timescales. Here we find comparable limits for those same FRBs analyzed by \cite{Tendulkar2016} with \textit{Swift} BAT. \\
\indent For comparison we also examine the raw rate data for FRB150215 and FRB160410\footnote{A flare from a nearby X-ray binary occurred at the same time as FRB110626 and so we exclude those results for the rate data here.}. We model the rate background emission over a total of 500 seconds as a linear fit in time. If necessary we use a low-order polynomial fit instead. We compute the root mean square of the background level in 200 second duration bins at $\pm$50 seconds from the time region of interest and denote the total number of counts in the bins as $N_{bins}$. We assume that the scatter within the region of interest also follows this scatter as well. From there we calculate 3$\sigma$ upper limits on the count rate assuming Poisson statistics. We then use XSPEC to convert the count rate limit to a fluence over the same energy range as the survey data. We find consistent results between the event rate data and survey images.\\
\begin{longtable*}{cccccccc}
\caption{3$\sigma$ upper limits to $\rm f_{\gamma}$ in different time ranges and energy bands for each FRB.} 
\\
 FRB Name & Bandpass & Date & Time & RA & Dec & $\rm \Delta t$ & $\rm f_{\gamma}$ \footnote{Values listed here are the fluence for the spectral template of a power law with an exponential cutoff ($E_\mathrm{peak}=1.5$ MeV). We also explore two softer spectral templates. See \S \ref{sec:GBM} for more information.}\\
 & &  [yyyy-mm-dd] & [hh:mm:ss] & & & [s] & [$10^{-6}$ $\ergcm$] \\ \hline \hline
 090625 & 8 keV - 40 MeV & 2009-06-25 & 21:53:51 & 46.95 & -29.93 & 100 & $<$7.9 \\
 &  &  &  &  &  & 10 & $<$2.5 \\
 &  &  &  &  &  & 1 & $<$0.82 \\
 &  &  &  &  &  & 0.1 & $<$0.28 \\ \hline
 110523 & 8 keV - 40 MeV & 2011-05-23 & 15:06:20 & 326.30 & -0.16 & 100 & $<$7.5 \\
 &  &  &  &  &  & 10 & $<$2.3 \\
 &  &  &  &  &  & 1 & $<$0.76 \\
 &  &  &  &  &  & 0.1 & $<$0.26 \\ \hline
 110626 & 8 keV - 40 MeV & 2011-06-26 & 21:33:16 & 315.75 & -44.73 & 100 & $<$7.5 \\
 &  &  &  &  &  & 10 & $<$2.3 \\
 &  &  &  &  &  & 1 & $<$0.76 \\
 &  &  &  &  &  & 0.1 & $<$0.26 \\ \hline
 110703 & 8 keV - 40 MeV & 2011-07-03 & 18:59:39 & 352.50 & -2.87 & 100 & $<$8.2 \\
 &  &  &  &  &  & 10 & $<$2.6 \\
 &  &  &  &  &  & 1 & $<$0.84 \\
 &  &  &  &  &  & 0.1 & $<$0.29 \\ \hline
 130628 & 8 keV - 40 MeV & 2013-06-28 & 03:57:59 & 135.76 & 3.44 & 100 & $<$6.6 \\
 &  &  &  &  &  & 10 & $<$2.1 \\
 &  &  &  &  &  & 1 & $<$0.7 \\
 &  &  &  &  &  & 0.1 & $<$0.24 \\ \hline
 130729 & 8 keV - 40 MeV & 2013-07-29 & 09:01:51 & 205.34 & -6.00 & 100 & $<$7.1 \\
 &  &  &  &  &  & 10 & $<$2.3 \\
 &  &  &  &  &  & 1 & $<$0.75 \\
 &  &  &  &  &  & 0.1 & $<$0.26 \\ \hline
 131104 & 8 keV - 40 MeV & 2013-11-04 & 18:04:00 & 101.04 & -51.28 & 100 & $<$8.4 \\
 &  &  &  &  &  & 10 & $<$2.7 \\
 &  &  &  &  &  & 1 & $<$0.87 \\
 &  &  &  &  &  & 0.1 & $<$0.3 \\ \hline
 150215 & 8 keV - 40 MeV & 2015-02-15 & 20:41:39 & 274.36 & -4.90 & 100 & $<$7.0 \\
 &  &  &  &  &  & 10 & $<$2.2 \\
 &  &  &  &  &  & 1 & $<$0.73 \\
 &  &  &  &  &  & 0.1 & $<$0.25 \\ \hline
 150418 & 8 keV - 40 MeV & 2015-04-18 & 04:29:05 & 109.12 & -19.04 & 100 & $<$7.1 \\
 &  &  &  &  &  & 10 & $<$2.3 \\
 &  &  &  &  &  & 1 & $<$0.74 \\
 &  &  &  &  &  & 0.1 & $<$0.25 \\ \hline
 150807 & 8 keV - 40 MeV & 2015-08-07 & 17:53:55 & 340.10 & -55.27 & 100 & $<$6.9 \\
 &  &  &  &  &  & 10 & $<$2.2 \\
 &  &  &  &  &  & 1 & $<$0.73 \\
 &  &  &  &  &  & 0.1 & $<$0.25 \\ \hline
 160317 & 8 keV - 40 MeV & 2016-03-17 & 09:00:30 & 118.45 & -29.61 & 100 & $<$7.0 \\
 &  &  &  &  &  & 10 & $<$2.2 \\
 &  &  &  &  &  & 1 & $<$0.73 \\
 &  &  &  &  &  & 0.1 & $<$0.25 \\ \hline
 160608 & 8 keV - 40 MeV & 2016-06-08 & 03:52:57 & 114.17 & -40.80 & 100 & $<$7.7 \\
 &  &  &  &  &  & 10 & $<$2.4 \\
 &  &  &  &  &  & 1 & $<$0.79 \\
 &  &  &  &  &  & 0.1 & $<$0.27 \\ \hline
 121102 3 & 8 keV - 40 MeV & 2015-05-17 & 17:51:41 & 82.99 & 33.15 & 100 & $<$7.6 \\
 &  &  &  &  &  & 10 & $<$2.4 \\
 &  &  &  &  &  & 1 & $<$0.78 \\
 &  &  &  &  &  & 0.1 & $<$0.26 \\ \hline
 121102 4 & 8 keV - 40 MeV & 2015-06-02 & 16:38:08 & 82.99 & 33.15 & 100 & $<$6.5 \\
 &  &  &  &  &  & 10 & $<$2.1 \\
 &  &  &  &  &  & 1 & $<$0.69 \\
 &  &  &  &  &  & 0.1 & $<$0.24 \\ \hline
 121102 5 & 8 keV - 40 MeV & 2015-06-02 & 16:47:36 & 82.99 & 33.15 & 100 & $<$6.7 \\
 &  &  &  &  &  & 10 & $<$2.1 \\
 &  &  &  &  &  & 1 & $<$0.69 \\
 &  &  &  &  &  & 0.1 & $<$0.24 \\ \hline
 121102 17 & 8 keV - 40 MeV & 2015-12-08 & 04:54:40 & 82.99 & 33.15 & 100 & $<$7.2 \\
 &  &  &  &  &  & 10 & $<$2.3 \\
 &  &  &  &  &  & 1 & $<$0.75 \\
 &  &  &  &  &  & 0.1 & $<$0.25 \\ \hline
 121102 18 & 8 keV - 40 MeV & 2016-08-23 & 17:51:24 & 82.99 & 33.15 & 100 & $<$7.6 \\
 &  &  &  &  &  & 10 & $<$2.4 \\
 &  &  &  &  &  & 1 & $<$0.79 \\
 &  &  &  &  &  & 0.1 & $<$0.26 \\ \hline
 121102 19 & 8 keV - 40 MeV & 2016-09-02 & 16:19:00 & 82.99 & 33.15 & 100 & $<$7.2 \\
 &  &  &  &  &  & 10 & $<$2.3 \\
 &  &  &  &  &  & 1 & $<$0.76 \\
 &  &  &  &  &  & 0.1 & $<$0.26 \\ \hline
 121102 20 & 8 keV - 40 MeV & 2016-09-02 & 16:41:02 & 82.99 & 33.15 & 100 & $<$7.7 \\
 &  &  &  &  &  & 10 & $<$2.4 \\
 &  &  &  &  &  & 1 & $<$0.79 \\
 &  &  &  &  &  & 0.1 & $<$0.27 \\ \hline
 121102 21 & 8 keV - 40 MeV & 2016-09-07 & 11:59:06 & 82.99 & 33.15 & 100 & $<$8 \\
 &  &  &  &  &  & 10 & $<$2.6 \\
 &  &  &  &  &  & 1 & $<$0.84 \\
 &  &  &  &  &  & 0.1 & $<$0.29 \\ \hline
 121102 22 & 8 keV - 40 MeV & 2016-09-12 & 10:58:31 & 82.99 & 33.15 & 100 & $<$7.3 \\
 &  &  &  &  &  & 10 & $<$2.4 \\
 &  &  &  &  &  & 1 & $<$0.78 \\
 &  &  &  &  &  & 0.1 & $<$0.26 \\ \hline
 121102 24 & 8 keV - 40 MeV & 2016-09-15 & 11:11:03 & 82.99 & 33.15 & 100 & $<$7.5 \\
 &  &  &  &  &  & 10 & $<$2.4 \\
 &  &  &  &  &  & 1 & $<$0.78 \\
 &  &  &  &  &  & 0.1 & $<$0.27 \\ \hline
 121102 27 & 8 keV - 40 MeV & 2016-09-17 & 10:29:09 & 82.99 & 33.15 & 100 & $<$7 \\
 &  &  &  &  &  & 10 & $<$2.2 \\
 &  &  &  &  &  & 1 & $<$0.73 \\
 &  &  &  &  &  & 0.1 & $<$0.25 \\ \hline
 121102 28 & 8 keV - 40 MeV & 2016-09-18 & 04:10:17 & 82.99 & 33.15 & 100 & $<$8 \\
 &  &  &  &  &  & 10 & $<$2.5 \\
 &  &  &  &  &  & 1 & $<$0.83 \\
 &  &  &  &  &  & 0.1 & $<$0.28 \\ \hline
 121102 29 & 8 keV - 40 MeV & 2016-09-18 & 05:14:14 & 82.99 & 33.15 & 100 & $<$8.5 \\
 &  &  &  &  &  & 10 & $<$2.7 \\
 &  &  &  &  &  & 1 & $<$0.89 \\
 &  &  &  &  &  & 0.1 & $<$0.3 \\ \hline
 121102 33 & 8 keV - 40 MeV & 2017-01-12 & 01:39:26 & 82.99 & 33.15 & 100 & $<$7.2 \\
 &  &  &  &  &  & 10 & $<$2.3 \\
 &  &  &  &  &  & 1 & $<$0.76 \\
 &  &  &  &  &  & 0.1 & $<$0.26 \\ \hline
 121102 34 & 8 keV - 40 MeV & 2017-01-12 & 02:25:12 & 82.99 & 33.15 & 100 & $<$6.6 \\
 &  &  &  &  &  & 10 & $<$2.1 \\
 &  &  &  &  &  & 1 & $<$0.69 \\
 &  &  &  &  &  & 0.1 & $<$0.24 \\ \hline
 121102 35 & 8 keV - 40 MeV & 2017-01-12 & 02:36:30 & 82.99 & 33.15 & 100 & $<$6.8 \\
 &  &  &  &  &  & 10 & $<$2.2 \\
 &  &  &  &  &  & 1 & $<$0.72 \\
 &  &  &  &  &  & 0.1 & $<$0.25 \\ \hline
 121102 37 & 8 keV - 40 MeV & 2017-01-12 & 03:16:33 & 82.99 & 33.15 & 100 & $<$6.8 \\
 &  &  &  &  &  & 10 & $<$2.2 \\
 &  &  &  &  &  & 1 & $<$0.71 \\
 &  &  &  &  &  & 0.1 & $<$0.24 \\ \hline
 121102 38 & 8 keV - 40 MeV & 2017-01-12 & 03:26:24 & 82.99 & 33.15 & 100 & $<$7.5 \\
 &  &  &  &  &  & 10 & $<$2.3 \\
 &  &  &  &  &  & 1 & $<$0.75 \\
 &  &  &  &  &  & 0.1 & $<$0.25 \\ \hline \hline
090625 & 60 MeV - 100 GeV & 2009-06-25 & 21:53:51 & 46.95 & -29.93 & 100 & $<$0.31 \\ 
 130628 & 60 MeV - 100 GeV & 2013-06-28 & 03:57:59 & 135.76 & 3.44 & 100 & $<$0.83 \\
 150215 & 60 MeV - 100 GeV & 2015-02-15 & 20:41:39 & 274.36 & -4.90 & 100 & $<$1.5 \\
 150418 & 60 MeV - 100 GeV & 2015-04-18 & 04:29:05 & 109.12 & -19.04 & 100 & $<$0.31 \\
 160317 & 60 MeV - 100 GeV & 2016-03-17 & 09:00:30 & 118.45 & -29.61 & 100 & $<$0.77 \\
 160608 & 60 MeV - 100 GeV & 2016-06-08 & 03:52:57 & 114.17 & -40.80 & 100 & $<$0.38 \\
 121102 18 & 60 MeV - 100 GeV & 2016-08-23 & 17:51:24 & 82.99 & 33.15 & 100 & $<$0.45 \\
 121102 19 & 60 MeV - 100 GeV & 2016-09-02 & 16:19:00 & 82.99 & 33.15 & 100 & $<$1.4 \\
 121102 22 & 60 MeV - 100 GeV & 2016-09-12 & 10:58:31 & 82.99 & 33.15 & 100 & $<$0.73 \\
 121102 27 & 60 MeV - 100 GeV & 2016-09-17 & 10:29:09 & 82.99 & 33.15 & 100 & $<$0.5 \\
 121102 34 & 60 MeV - 100 GeV & 2017-01-12 & 02:25:12 & 82.99 & 33.15 & 100 & $<$1.4 \\ \hline \hline
 110626 & 15-350 keV & 2011-06-26 & 21:33:16 & 315.75 & -44.73 & 300 & $<$4.8 \\ \hline
150215 & 15-350 keV & 2015-02-15 & 20:41:39 & 274.36 & -4.90 & 300 & $<$2.3 \\
 &  &  &  &  &  & 100 & $<$0.065 \\
 &  &  &  &  &  & 10 & $<$0.055 \\
 &  &  &  &  &  & 1 & $<$0.027 \\
 &  &  &  &  &  & 0.064 & $<$0.0092 \\ \hline
 160410 & 15-350 keV & 2016-04-10 & 08:33:38 & 130.35 & 6.08 & 300 & $<$1.6 \\
 &  &  &  &  &  & 100 & $<$0.17\\
 &  &  &  &  &  & 10 & $<$0.08 \\
 &  &  &  &  &  & 1 & $<$0.021 \\
 &  &  &  &  &  & 0.064 & $<$0.0048 \\ \hline \hline
\label{tablat}

\end{longtable*}

\section{Analysis and Interpretation} \label{sec:analysis}
\indent We report no significant excess in high-energy emission from the \textit{Fermi} GBM, \textit{Fermi} LAT, or \textit{Swift} BAT for any of the individual FRBs or repeats from FRB 121102. The expected high-energy fluence from FRBs is highly model-dependent. Given the number of theories in the literature we take a two-fold approach in this work. First, we compare our results with previously reported observations of high-energy counterparts to FRBs, such as that claimed for FRB131104. Second, we consider the implications of our non-detections for some of the more plausible models that have been considered. \\

\subsection{Limits on the ratio of radio to gamma-ray fluence \label{sec:ratiolim}}
\indent A recent paper by DeLaunay et al. (2016) reports a possible connection of FRB131104 to a \textit{Swift} BAT long GRB with fluence $f_{\gamma} \approx 4 \times 10^{-6} \ergcm$ and duration $T_{90} = 377$ s, where $T_{90}$ is defined as the time over which a burst emits from 5\% of its total measured counts to 95\%. With the reported radio fluence for FRB131104 of 2.33 \mbox{Jy ms}, this implies a ratio of radio to gamma-ray emission of $\approx6 \times 10^{5} \ \rat$. For consistency with the DeLaunay result we consider fluences on 100 s timescales. The radio fluences are taken from the FRBCAT. We find that $\nicefrac{f_{r}}{f_{\gamma}} \gtrsim 10^5-10^7 \ \rat $ for the non-repeating FRBs in our sample and we find $\nicefrac{f_{r}}{f_{\gamma}} \gtrsim 10^4-10^5 \ \rat $ for the repeating bursts of FRB121102 (Table \ref{tab:ratios}). None of the limits derived from the \textit{Swift} BAT or \textit{Fermi} LAT are consistent with the DeLaunay result, providing lower limits to the radio to gamma-ray emission ratio that exceed their reported values. However nine out of the eleven non-repeating FRBs and all of the repeating bursts from the \textit{Fermi} GBM are consistent.\\ 
\indent We then compare our limits to those expected from magnetar hyperflares (see $\S$\ref{sec:mag}), given observations of SGR 1806$-$20. \cite{Tendulkar2016} find upper limits of $\nicefrac{f_{r}}{f_{\gamma}} < 10^7 \ \rat $ for the giant flare event on December 27, 2004, based on archival observations of FRBs taken with the Konus-Wind gamma-ray spectrometer, the \textit{Swift} BAT, and the \textit{Fermi} GBM. Although the timescales and the bandpass of the Konus-Wind (10 keV to 10 MeV) are not identical to our analysis, this is inconsistent with four of the non-repeating FRBs yet consistent with all of the repeating bursts (10 in our sample) for limits on timescales of order 0.1 s (Table \ref{tab:ratios}). \\
\indent We also compare our ratios to those of \cite{Scholz2017b} who find a lower limit on $\nicefrac{f_{r}}{f_{\gamma}}$ of $> 2 \times 10^{8}$ based on \textit{Fermi} GBM observations of the Repeater. Although they look at bursts on finer timescales of a few hundred milliseconds this is consistent with all of our \textit{Fermi} GBM limits on timescales of 0.1 s.\\

\begin{table*}[]
\centering
\caption{Maximum 3$\sigma$ ratio of radio to high-energy emission on 100 s and  0.1 s timescales.}
\begin{tabular}{cccccccc}
  FRB Name & Bandpass & $\rm \Delta t$& $f_{r}$  & $f_{\gamma}$ & log$(\nicefrac{f_{r}}{f_{\gamma}})$  \\ 
  & & [s] & [$\rm Jy \ ms$]&[$10^{-6}$$\ergcm$] & [$\rm Jy \ ms \ erg^{-1} \ cm^{2}$]&   \\ \hline \hline
090625 & 8-4e4 keV & 100 & 2.19 & $<$7.9 & $>$5.44 \\
 & & 0.1 &  & $<$0.28 & $>$6.89 \\ \hline
 110523 & 8-4e4 keV & 100 & 1.04 & $<$7.5 & $>$5.14 \\
 & & 0.1 &  & $<$0.26 & $>$6.60 \\ \hline
 110626 & 8-4e4 keV & 100 & 0.56 & $<$7.5 & $>$4.88 \\
 & & 0.1 &  & $<$0.26 & $>$6.33 \\ \hline
 110703 & 8-4e4 keV & 100 & 1.80 & $<$8.2 & $>$5.34 \\
 & & 0.1 &  & $<$0.29 & $>$6.80 \\ \hline
 130628 & 8-4e4 keV & 100 & 1.22 & $<$6.6 & $>$5.26 \\
 & & 0.1 &  & $<$0.24 & $>$6.70 \\ \hline
 130729 & 8-4e4 keV & 100 & 3.43 & $<$7.1 & $>$5.68 \\
 & & 0.1 &  & $<$0.26 & $>$7.13 \\ \hline
 131104 & 8-4e4 keV & 100 & 2.33 & $<$8.4 & $>$5.45 \\
 & & 0.1 &  & $<$0.3 & $>$6.89 \\ \hline
 150215 & 8-4e4 keV & 100 & 1.96 & $<$7 & $>$5.45 \\
 & & 0.1 &  & $<$0.25 & $>$6.90 \\ \hline
 150418 & 8-4e4 keV & 100 & 1.76 & $<$7.1 & $>$5.39 \\
 & & 0.1 &  & $<$0.25 & $>$6.84 \\ \hline
 150807 & 8-4e4 keV & 100 & 44.80 & $<$6.9 & $>$6.81 \\
 & & 0.1 &  & $<$0.25 & $>$8.26 \\ \hline
 160317 & 8-4e4 keV & 100 & 69.00 & $<$7 & $>$7.00 \\
 & & 0.1 &  & $<$0.25 & $>$8.44 \\ \hline
 160608 & 8-4e4 keV & 100 & 37.00 & $<$7.7 & $>$6.68 \\
 & & 0.1 &  & $<$0.27 & $>$8.14 \\ \hline
 121102 3 & 8-4e4 keV & 100 & 0.10 & $<$7.6 & $>$4.12 \\
 & & 0.1 &  & $<$0.26 & $>$5.58 \\ \hline
 121102 4 & 8-4e4 keV & 100 & 0.20 & $<$6.5 & $>$4.49 \\
 & & 0.1 &  & $<$0.24 & $>$5.93 \\ \hline
 121102 5 & 8-4e4 keV & 100 & 0.09 & $<$6.7 & $>$4.13 \\
 & & 0.1 &  & $<$0.24 & $>$5.58 \\ \hline
 121102 17 & 8-4e4 keV & 100 & 0.09 & $<$7.2 & $>$4.10 \\
 & & 0.1 &  & $<$0.25 & $>$5.55 \\ \hline
 121102 28 & 8-4e4 keV & 100 & 0.36 & $<$8 & $>$4.65 \\
 & & 0.1 &  & $<$0.28 & $>$6.11 \\ \hline
 121102 29 & 8-4e4 keV & 100 & 0.29 & $<$8.5 & $>$4.53 \\
 & & 0.1 &  & $<$0.3 & $>$5.98 \\ \hline
 121102 33 & 8-4e4 keV & 100 & 0.62 & $<$7.2 & $>$4.93 \\
 & & 0.1 &  & $<$0.26 & $>$6.38 \\ \hline
 121102 35 & 8-4e4 keV & 100 & 0.03 & $<$6.8 & $>$3.65 \\
 & & 0.1 &  & $<$0.25 & $>$5.09 \\ \hline
 121102 37 & 8-4e4 keV & 100 & 0.22 & $<$6.8 & $>$4.51 \\
 & & 0.1 &  & $<$0.24 & $>$5.96 \\ \hline
 121102 38 & 8-4e4 keV & 100 & 0.10 & $<$7.5 & $>$4.13 \\
 & & 0.1 &  & $<$0.25 & $>$5.60 \\ \hline \hline
090625 & 60-1e5 MeV & 100 & 2.19 & $<$0.31 & $>$6.85 \\
 130628 & 60-1e5 MeV & 100 & 1.22 & $<$0.83 & $>$6.17 \\
 150215 & 60-1e5 MeV & 100 & 1.96 & $<$1.5 & $>$6.10 \\
 150418 & 60-1e5 MeV & 100 & 1.76 & $<$0.31 & $>$6.75 \\
 160317 & 60-1e5 MeV & 100 & 69.00 & $<$0.77 & $>$7.95 \\
 160608 & 60-1e5 MeV & 100 & 37.00 & $<$0.38 & $>$7.99 \\ \hline \hline
150215 & 15-350 keV & 100 & 1.96 & $<$0.065 & $>$7.48 \\
 & & 0.1 &  & $<$0.0092 & $>$5.60 \\ \hline
 160410 & 15-350 keV & 100 & 34.00 & $<$0.17 & $>$8.29 \\
 & & 0.1 &  & $<$0.0048 & $>$9.85 \\ \hline \hline
\end{tabular}
\label{tab:ratios}
\end{table*}

\subsection{Constraints on theoretical models}
\indent Given that there are dozens of theories put forth attempting to explain FRBs, because of their implied small sizes ($<$300 km), we choose to favor models involving compact objects like neutron stars. We cannot examine all models in our analysis here so we consider only those models which satisfy the criteria of a compact emission region, extragalactic distance scale, coherent emission mechanism, repeated outbursts from at least some FRBs, and large all-sky rates. Here we consider the ramifications of our results in reference to some of the more probable theories. \\

\subsubsection{Rotationally powered pulses from neutron stars}
\indent The Crab pulsar exhibits rare, giant radio pulse behavior at GHz frequencies. Giant pulse occurrences are random in time but are correlated with the pulsar's main pulse or interpulse periods. About 1\% of pulses from the Crab are giant pulses. These giant bursts can exceed 0.5 MJy over a duration of a few nanoseconds \citep{CordesWasserman2016}. The most extreme event was a 0.4 ns pulse with a flux density of 2.2 MJy at 9 GHz \citep{Hankins2007}. The short durations, large fluxes, and non-periodic nature of giant pulses make them an excellent test for comparison with FRBs.\\
\indent The Crab emits across all frequencies and also exhibits giant pulse behavior in the gamma- and X-ray as well \citep{Buhler2014}. \cite{Mickaliger2012} examine the correlation between radio giant pulses and high-energy photons from 0.1 to 100 GeV and find no significant association. \\
\indent If FRBs are powered by giant pulses from pulsars then we would expect them to be nearby. \cite{CordesWasserman2016} show that even the most extreme giant pulse from the Crab could not provide the necessary radio fluences of $\sim$2 Jy ms at 1 Gpc (the reported distance of the Repeater). For these observed fluences they find a maximum distance for Crab-like giant pulses of $\sim$100 Mpc. \\
\indent An extremely energetic burst of gamma rays from the Crab occurred in April 2011 with a luminosity of \mbox{$\rm L_{\gamma} = 4 \times 10^{36} \ erg \ s^{-1}$} \citep{Striani2011,Buehler2012}. If we assume this energy scale for FRBs, then at 1 Gpc we would expect to observe a fluence around 20 orders of magnitude fainter than the background level. The expected flux density is orders of magnitude too small at 1 Gpc to account for the radio emission observed. However, if indeed the FRBs are located at these Galactic distances this would imply that most of the DM is local to the source, rather than from the IGM. Although the distances derived here are inconsistent with that of the Repeater we still consider giant Crab-like pulses as a viable model for FRBs since the energy scale and lack of high-energy emission are consistent with that observed for FRBs. \\

\subsubsection{Magnetically powered pulses from neutron stars \label{sec:mag}}
\indent Magnetars are highly-magnetized neutron stars with surface magnetic field strengths of \mbox{$ \rm B_{surf} \sim 10^{14}$ - $10^{15}$ G} \citep{Duncan1992}. They are known to regularly emit hard X-ray/soft gamma-ray flares of duration $<$1s with total energy \mbox{$10^{41}$ erg} \citep{Kouveliotou1998}. Distinct from these ``average flares'', magnetars can also emit hyperflares which are several orders of magnitude higher in energy. A hyperflare is marked by a millisecond rise time, hard X-ray peak, and an oscillating tail lasting for minutes. Although there are $\sim$30 magnetars known to date, there are only three observed hyperflare events, with the SGR 1806$-$20 event being the most energetic \citep{Mazets1979, Hurley1999, Hurley2005, Palmer2005}. \\
\indent Magnetar hyperflares are a popular theory for FRBs \citep{Popov2013, Kulkarni2015, Pen2015, Lyubarsky2014}. They have sub-second time variation, extreme energetics, and (depending on assumptions made about the underlying magnetar population) comparable event rates \citep{Nicholl2017}.  Magnetars are thought to be correlated with recent star formation and should therefore be enshrouded in dense gas and dust. This would imply a significant portion of an FRB's DM can be attributed to its local environment, rather than to the IGM, placing them at extragalactic, but not necessarily cosmological, distances. The properties of the host galaxy of the repeating FRB (e.g., a low mass, low metallicity, star-forming dwarf galaxy at redshift $z\sim$0.2) is consistent with where we might expect to find magnetars \citep{Tendulkar2017}.  \\
\indent If FRBs are caused by magnetar-like hyperflares we can place constraints on their distances by assuming a similar energy release to the SGR 1806$-$20 event. The total flare energy of this event is $10^{47}$ erg and could only have been observed out to 40 Mpc before falling below the threshold of most X-ray/gamma-ray telescopes. Figure \ref{fig:dist_plots} shows the inferred $E_{\gamma}$ for each FRB at different distances, neglecting k-corrections\footnote{K-corrections allow a conversion between a measurement at a redshift, $z$, to its equivalent rest-frame value.}. We can be conservative and consider possibilities that other hyperflares could be stronger than the SGR 1806$-$20 event, therefore pushing the maximum energy release to $E_{\gamma} = 10^{49}$ erg. For fluence upper limits set by the harder spectral template in GBM we find the FRBs should be located no nearer than about 0.5 Gpc, which is consistent with the distance of the Repeater at 1 Gpc. \\
\indent There are a few caveats with this picture. One is that, based on the SGR1806$-$20 event, we may not expect to see any radio emission from these hyperflares (see $\S$\ref{sec:ratiolim}). On the other hand, \cite{Lyutikov2002} proposes a model where radio counterparts could be seen at \mbox{$\sim$1 Gpc} distances. We note that we are extrapolating the properties of all giant magnetar flares from a total sample of three and that it is impossible to yet know what other subclasses of magnetar hyperflares might actually exist. Another issue with the theory is that, based on constraints from the DM, optical depth, and expansion of the supernova (SN) ejecta surrounding the magnetar, the age of the source must be less than 100 years \citep{Metzger2017}. This implies that as the remnant expands in time we should expect to see the observed DM evolve as well, despite there being no such evidence for this based on the DM of the repeating bursts from FRB121102. However, given these caveats we still consider the theory that FRBs originate as magnetar hyperflares as plausible.  \\

\subsubsection{Coalescence models \label{sec:coal}}
\begin{figure}[h] 
     \begin{center}
        \subfigure[]{%
            \includegraphics[width=0.5\textwidth]{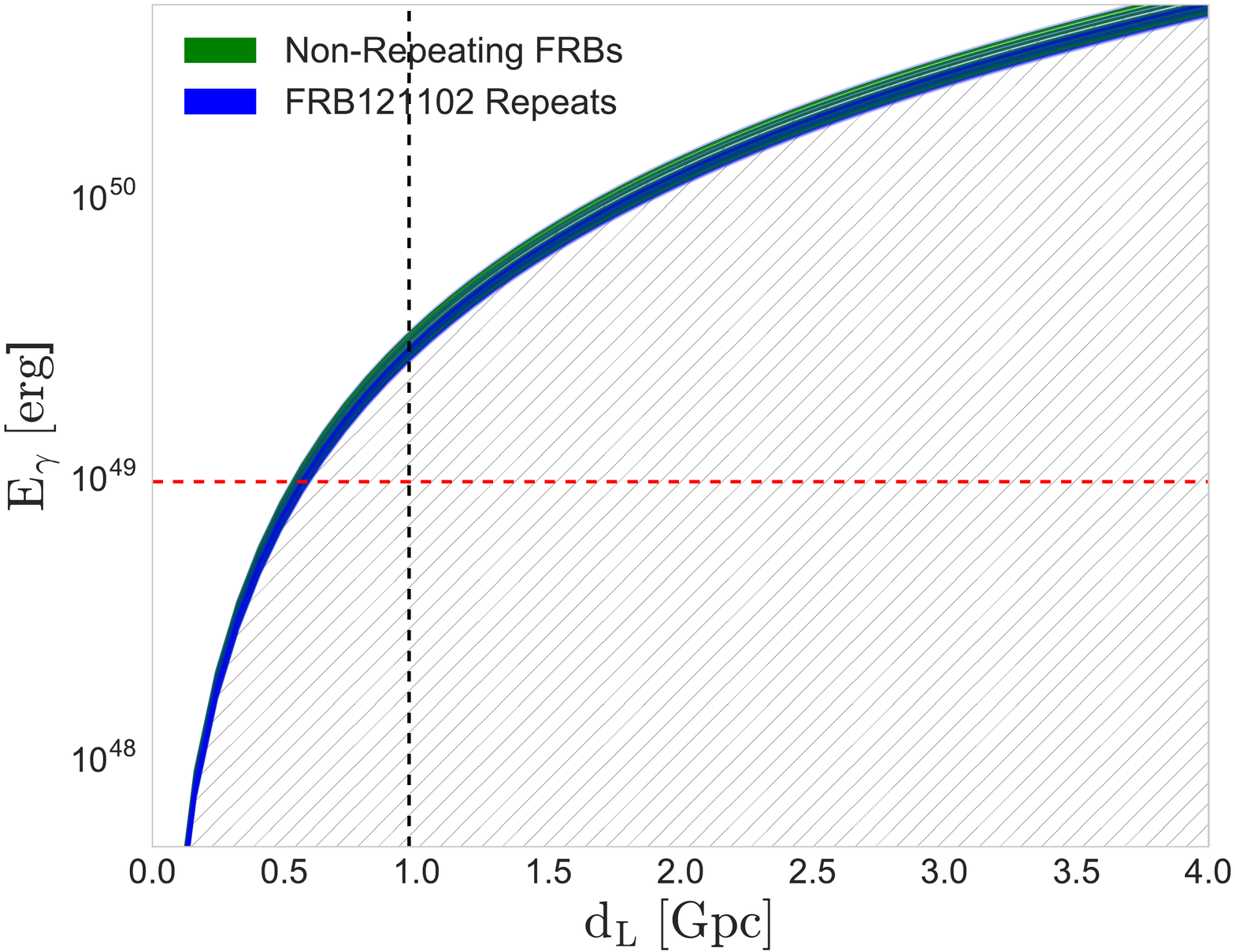}
        }\\
        \subfigure[]{%
            \includegraphics[width=0.5\textwidth]{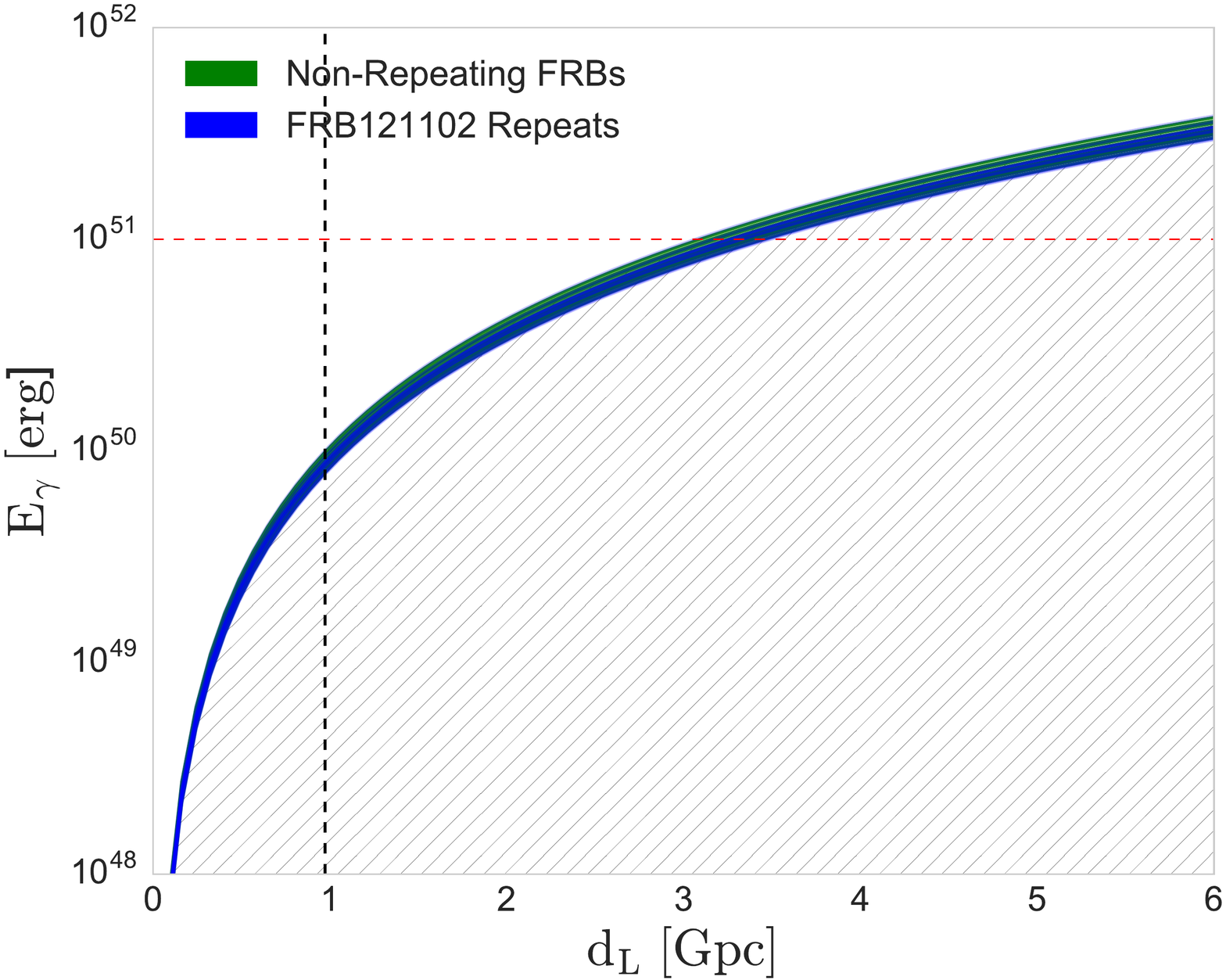}
        }%
    \end{center}
    \caption{Given the fluence upper limits on varying timescales (Table \ref{tablat}) we can predict how far away FRBs can be observed as predicted by different models. The hatched pattern represents viable parameter space. The black dashed line is the reported distance of the host galaxy of FRB121102. a.) The red dashed line represents the energy cut-off for magnetar hyperflares ($E_{\gamma} < 10^{49}$ erg). b.) The red dashed line represents the energy cut-off for coalesence models ($E_{\gamma} < 10^{51}$ erg). }
    \label{fig:dist_plots}
\end{figure}

\indent \cite{Dokuchaev2017} propose a model where FRBs are caused by collisions between neutron stars in the centers of evolved galaxies. This coalescence is suggested to generate short GRBs and the extreme energies produced have led some to suggest they could also power FRBs \citep{Takami2014, Berger2014}. The model predicts that a binary merger is not necessary to generate an FRB signal. As the neutron stars inspiral, their magnetic fields become synchronized with the binary rotation. This can result in magnetic reconnections which produce coherent radio emission.\\
\indent The inferred rate of FRBs is much higher than that of neutron star - neutron star mergers. Only the most optimistic binary neutron star merger rates could begin to compare with the lowest expected FRB rates \citep{Callister2016}. Assuming this is the case however would imply that the majority of binary neutron star mergers will result in an observable FRB. This is in apparent conflict with the small number of known FRBs - only 70 to date (FRBCAT) - and the lack of any associated FRB with the recent GRB170817A/GW170817 event \citep{LVCMMAPaper}.  However we caution that radio observations of GW170817 did not begin until approximately 12 hours after the merger, making firm conclusions difficult to draw. \\
\indent With the notable exception of 
GRB170817A (a highly sub-luminous event;  \citealt{LVC+GBMPaper}), the prompt isotropic 
energy release of short GRBs is $\sim 10^{51}-10^{52}$ 
erg \citep{Berger2014}. 
The divide between short and long GRBs occurs at about 2 s \citep{Kouveliotou1993}, so therefore we consider time scales of high-energy emission of 1 s. We take a similar approach to the magnetar model and constrain distances out to which we should expect to see FRBs if they are powered by coalescence (Figure \ref{fig:dist_plots}). We find that colliding neutron stars must reside outside of $\sim3$ Gpc to account for the lack of detected high-energy emission we observe.  This is inconsistent with the FRB121102 result and the observed DMs of the other FRBs. If we attribute all of the observed DM to propagation through the IGM then we would expect FRBs to reside at distances of no more than $\sim$3 Gpc \citep{Thornton2013}. In addition, low-luminosity, GRB170817A-like events must be located at distances further than a few hundred Mpc to account for a lack of radio emission. This is inconsistent with the merger's reported distance of only 40 Mpc \citep{LVCMMAPaper}. \\
\indent If FRBs are the result of collisions between neutron stars then the absence of gamma-ray emission is puzzling. \citet{Dokuchaev2017} propose that the GRB occurred off-axis and we are left seeing only the radio afterglow. In addition, they propose that collisions of this kind may also produce relativistic fireballs which can be lensed by the central supermassive black hole. The effects of this lensing is to produce a range of achromatic flashes of varying wavelengths. In this case the gas produced by the collision could absorb some of the high-energy emission. However, given the inconsistencies between this model and the FRB rate, observed FRB DMs, and the distances of the Repeater and GRB170817A we consider it unlikely that FRBs are caused by binary neutron star mergers. \\

\subsubsection{``Cosmic combs''}
\indent \cite{Zhang2017} proposes a model that can reproduce the variety of observations associated with FRBs (e.g., the gamma-ray signal associated with FRB131104, the active galactic nucleus (AGN) possibly coincident with FRB150418\footnote{\cite{Keane2016} use the coincidence of FRB150418 to a fading radio transient to identify a host galaxy at $z \sim 0.5$. \cite{Williams2016} claim that the radio source is instead AGN variability and is not connected to the FRB.}, and the repeating nature of FRB121102). The magnetosphere of a cosmological pulsar can be ``combed'' by a passing astrophysical plasma stream and accelerated by magnetic reconnections to produce a FRB. The origins of the plasma stream will determine what signatures are detected. For example, Zhang proposes that the radio flare associated with the FRB150418 event is in fact the original plasma stream which combed a pulsar to create the observed FRB. Also, they suggest that the repeater could be powered by irregular emission from a supernova remnant. Anything from AGN, GRBs, SNe, tidal disruption events, or stellar flares could be responsible for combing these signals from pulsars. The only condition needed to produce such a phenomenon is that the ram pressure of the plasma stream from these objects exceeds the magnetic pressure of the magnetosphere of the pulsar.\\
\indent Similar to $\S$\ref{sec:coal}, we consider the proposition that FRBs are caused by short GRBs originating from binary neutron star collisions. If we consider FRBs as counterparts to combed GRB signals then we can use a statistical approach to determine the maximum percentage of events that must come from GRB-like sources in order to account for the observed high-energy non-detections. We assume a binomial distribution of cosmic comb outcomes where p is the probability of a cosmic comb event originating from a GRB and take our sample size to be the n = 12 non-repeating FRBs in Table \ref{tablat}. Therefore the probability of getting k observed high-energy events is: \begin{equation}
    P = \binom{n}{k} p^k (1-p)^{n-k}. 
\end{equation} Since we report no significant high-energy counterparts, the probability of getting k=0 events is $<$17.5\% (at 90\% confidence).  If instead we calculate the probability of finding k=0 events over all observations (both repeating and non-repeating, n = 30) then this decreases to $<$7.4\%. Therefore we disfavor GRB cosmic combs as a plausible explanation for the origins of FRBs. \\ 

\subsubsection{Other Compact Object Models}
\indent To date, there exist dozens of theories in the literature describing FRB origins. In this section we summarize additional models involving compact objects which we feel do not warrant the full analytical treatment exhibited in previous sections.  \\
\indent FRBs could be produced by collapsing supramassive neutron stars \citep{Falke2014,Zhang2014}. While the timescale of collapse is consistent with that of FRBs, it fails to explain any repeating phenomena or the production of the radio emission itself. Similarly to $\S$\ref{sec:coal}, binaries involving neutron stars and white dwarfs have been proposed \citep{Gu2016,Lin2018} although are specifically invoked to explain only the Repeater. Several models exist involving neutron star interactions with black holes \citep{Abramowicz2018,Bhatt2017}, black hole interactions with white dwarfs \citep{Li2018}, and events from various other types of black holes and AGN \citep{Vieyro2017,Zhang2016}, but these theories all remain highly speculative.\\
\indent For a full treatment of all plausible theories on FRB origins we direct the reader to recent reviews which cover models involving both compact and non-compact sources \citep{Lorimer2018,Katz2018}.\\

\section{Conclusions} \label{sec:conclusions}
\indent We searched for high-energy counterparts to FRBs in \textit{Fermi} GBM, \textit{Fermi} LAT, and \textit{Swift} BAT. We detect no significant high-energy emission on timescales of several 0.1 to 100 s. We report upper limits to the emission in Table \ref{tablat} for each timescale (0.1, 1, 10, and 100 s) and energy range (15-350 keV, 300-40,000 keV, and 60-100,000 MeV) and also report limits on the ratio of radio to high-energy fluence for timescales of 0.1 and 100 s (Table \ref{tab:ratios}). \\
\indent We consider the implications of non-detections in the context of several theoretical models. We regard the neutron star coalescence model as highly unlikely as it is inconsistent with the observed FRB DMs, the number of observed FRBs to date, and the distance of the FRB121102 host galaxy. In addition, if the cosmic comb model explains FRBs then it is unlikely that FRBs are caused by GRBs ``combing" pulsars. \\
\indent Two of the more promising theories - magnetically or rotationally powered neutron stars - remain viable. We place lower limits on the distance for magnetar hyperflares which are consistent with the observed FRB DMs and the FRB121102 result. While the non-detection of high-energy emission agrees with the rotationally powered theory, it does not agree with the distance of the repeater. \\
\indent Although we exclude FRB131104 due to its low partial coding fraction we compare our results from the other FRBs with that of its claimed counterpart in BAT \citep{DeLaunay2016}. If FRBs are caused by similar events as that reported by DeLaunay then for the majority of our sample the observed gamma-ray fluence should have been larger than reported here. \\ 
\indent As this paper was being written new FRBs have been reported, including the second repeating burst FRB180814.J0422+73 (see the FRB Catalog at \url{frbcat.org} \citep{Petroff2016}). We will continue to explore the high-energy properties of these (and any other future FRBs) with the same methods as described in this paper. However for best results a dedicated, multi-wavelength follow-up procedure needs to be put in place. Ideally there would exist a joint campaign between telescopes for co-observing candidates so that data at other wavelengths would be immediately available. If there are in fact no counterparts to FRBs at other wavelengths, then future progress in the field will require precise localization from radio measurements, in particular interferometry. \\

\begin{acknowledgments}
The \textit{Fermi} LAT Collaboration acknowledges generous ongoing support
from a number of agencies and institutes that have supported both the
development and the operation of the LAT as well as scientific data analysis.
These include the National Aeronautics and Space Administration and the
Department of Energy in the United States, the Commissariat \`a l'Energie Atomique
and the Centre National de la Recherche Scientifique / Institut National de Physique
Nucl\'eaire et de Physique des Particules in France, the Agenzia Spaziale Italiana
and the Istituto Nazionale di Fisica Nucleare in Italy, the Ministry of Education,
Culture, Sports, Science and Technology (MEXT), High Energy Accelerator Research
Organization (KEK) and Japan Aerospace Exploration Agency (JAXA) in Japan, and
the K.~A.~Wallenberg Foundation, the Swedish Research Council and the
Swedish National Space Board in Sweden.
 
Additional support for science analysis during the operations phase is gratefully
acknowledged from the Istituto Nazionale di Astrofisica in Italy and the Centre
National d'\'Etudes Spatiales in France. This work performed in part under DOE
Contract DE-AC02-76SF00515.
\end{acknowledgments}

\bibliographystyle{aasjournal} 
\bibliography{ref} 

\begin{thebibliography}{}
\expandafter\ifx\csname natexlab\endcsname\relax\def\natexlab#1{#1}\fi
\providecommand{\url}[1]{\href{#1}{#1}}

\bibitem[{{Abbott} {et~al.}(2017{\natexlab{a}}){Abbott}, {Abbott}, {Abbott},
  {Acernese}, {Ackley}, {Adams}, {Adams}, {Addesso}, {Adhikari}, {Adya}, \&
  et~al.}]{LVCMMAPaper}
{Abbott}, B.~P., {Abbott}, R., {Abbott}, T.~D., {et~al.} 2017{\natexlab{a}},
  \apjl, 848, L12

\bibitem[{{Abbott} {et~al.}(2017{\natexlab{b}}){Abbott}, {Abbott}, {Abbott},
  {Acernese}, {Ackley}, {Adams}, {Adams}, {Addesso}, {Adhikari}, {Adya}, \&
  et~al.}]{LVC+GBMPaper}
---. 2017{\natexlab{b}}, \apjl, 848, L13

\bibitem[{{Abramowicz} {et~al.}(2018){Abramowicz}, {Bejger}, \&
  {Wielgus}}]{Abramowicz2018}
{Abramowicz}, M.~A., {Bejger}, M., \& {Wielgus}, M. 2018, \apj, 868, 17

\bibitem[{{Acero} {et~al.}(2015){Acero}, {Ackermann}, {Ajello}, {Albert},
  {Atwood}, {Axelsson}, {Baldini}, {Ballet}, {Barbiellini}, {Bastieri},
  {Belfiore}, {Bellazzini}, {Bissaldi}, {Blandford}, {Bloom}, {Bogart},
  {Bonino}, {Bottacini}, {Bregeon}, {Britto}, {Bruel}, {Buehler}, {Burnett},
  {Buson}, {Caliandro}, {Cameron}, {Caputo}, {Caragiulo}, {Caraveo},
  {Casandjian}, {Cavazzuti}, {Charles}, {Chaves}, {Chekhtman}, {Cheung},
  {Chiang}, {Chiaro}, {Ciprini}, {Claus}, {Cohen-Tanugi}, {Cominsky}, {Conrad},
  {Cutini}, {D'Ammando}, {de Angelis}, {DeKlotz}, {de Palma}, {Desiante},
  {Digel}, {Di Venere}, {Drell}, {Dubois}, {Dumora}, {Favuzzi}, {Fegan},
  {Ferrara}, {Finke}, {Franckowiak}, {Fukazawa}, {Funk}, {Fusco}, {Gargano},
  {Gasparrini}, {Giebels}, {Giglietto}, {Giommi}, {Giordano}, {Giroletti},
  {Glanzman}, {Godfrey}, {Grenier}, {Grondin}, {Grove}, {Guillemot}, {Guiriec},
  {Hadasch}, {Harding}, {Hays}, {Hewitt}, {Hill}, {Horan}, {Iafrate}, {Jogler},
  {J{\'o}hannesson}, {Johnson}, {Johnson}, {Johnson}, {Johnson}, {Kamae},
  {Kataoka}, {Katsuta}, {Kuss}, {La Mura}, {Landriu}, {Larsson}, {Latronico},
  {Lemoine-Goumard}, {Li}, {Li}, {Longo}, {Loparco}, {Lott}, {Lovellette},
  {Lubrano}, {Madejski}, {Massaro}, {Mayer}, {Mazziotta}, {McEnery},
  {Michelson}, {Mirabal}, {Mizuno}, {Moiseev}, {Mongelli}, {Monzani},
  {Morselli}, {Moskalenko}, {Murgia}, {Nuss}, {Ohno}, {Ohsugi}, {Omodei},
  {Orienti}, {Orlando}, {Ormes}, {Paneque}, {Panetta}, {Perkins},
  {Pesce-Rollins}, {Piron}, {Pivato}, {Porter}, {Racusin}, {Rando}, {Razzano},
  {Razzaque}, {Reimer}, {Reimer}, {Reposeur}, {Rochester}, {Romani},
  {Salvetti}, {S{\'a}nchez-Conde}, {Saz Parkinson}, {Schulz}, {Siskind},
  {Smith}, {Spada}, {Spandre}, {Spinelli}, {Stephens}, {Strong}, {Suson},
  {Takahashi}, {Takahashi}, {Tanaka}, {Thayer}, {Thayer}, {Thompson},
  {Tibaldo}, {Tibolla}, {Torres}, {Torresi}, {Tosti}, {Troja}, {Van Klaveren},
  {Vianello}, {Winer}, {Wood}, {Wood}, {Zimmer}, \& {Fermi-LAT
  Collaboration}}]{Acero2015}
{Acero}, F., {Ackermann}, M., {Ajello}, M., {et~al.} 2015, \apjs, 218, 23

\bibitem[{{Ackermann} {et~al.}(2012){Ackermann}, {Ajello}, {Albert},
  {Allafort}, {Atwood}, {Axelsson}, {Baldini}, {Ballet}, {Barbiellini},
  {Bastieri}, {Bechtol}, {Bellazzini}, {Bissaldi}, {Blandford}, {Bloom},
  {Bogart}, {Bonamente}, {Borgland}, {Bottacini}, {Bouvier}, {Brandt},
  {Bregeon}, {Brigida}, {Bruel}, {Buehler}, {Burnett}, {Buson}, {Caliandro},
  {Cameron}, {Caraveo}, {Casandjian}, {Cavazzuti}, {Cecchi}, {{\c C}elik},
  {Charles}, {Chaves}, {Chekhtman}, {Cheung}, {Chiang}, {Ciprini}, {Claus},
  {Cohen-Tanugi}, {Conrad}, {Corbet}, {Cutini}, {D'Ammando}, {Davis}, {de
  Angelis}, {DeKlotz}, {de Palma}, {Dermer}, {Digel}, {Silva}, {Drell},
  {Drlica-Wagner}, {Dubois}, {Favuzzi}, {Fegan}, {Ferrara}, {Focke}, {Fortin},
  {Fukazawa}, {Funk}, {Fusco}, {Gargano}, {Gasparrini}, {Gehrels}, {Giebels},
  {Giglietto}, {Giordano}, {Giroletti}, {Glanzman}, {Godfrey}, {Grenier},
  {Grove}, {Guiriec}, {Hadasch}, {Hayashida}, {Hays}, {Horan}, {Hou}, {Hughes},
  {Jackson}, {Jogler}, {J{\'o}hannesson}, {Johnson}, {Johnson}, {Johnson},
  {Kamae}, {Katagiri}, {Kataoka}, {Kerr}, {Kn{\"o}dlseder}, {Kuss}, {Lande},
  {Larsson}, {Latronico}, {Lavalley}, {Lemoine-Goumard}, {Longo}, {Loparco},
  {Lott}, {Lovellette}, {Lubrano}, {Mazziotta}, {McConville}, {McEnery},
  {Mehault}, {Michelson}, {Mitthumsiri}, {Mizuno}, {Moiseev}, {Monte},
  {Monzani}, {Morselli}, {Moskalenko}, {Murgia}, {Naumann-Godo}, {Nemmen},
  {Nishino}, {Norris}, {Nuss}, {Ohno}, {Ohsugi}, {Okumura}, {Omodei},
  {Orienti}, {Orlando}, {Ormes}, {Paneque}, {Panetta}, {Perkins},
  {Pesce-Rollins}, {Pierbattista}, {Piron}, {Pivato}, {Porter}, {Racusin},
  {Rain{\`o}}, {Rando}, {Razzano}, {Razzaque}, {Reimer}, {Reimer}, {Reposeur},
  {Reyes}, {Ritz}, {Rochester}, {Romoli}, {Roth}, {Sadrozinski}, {Sanchez},
  {Saz Parkinson}, {Sbarra}, {Scargle}, {Sgr{\`o}}, {Siegal-Gaskins},
  {Siskind}, {Spandre}, {Spinelli}, {Stephens}, {Suson}, {Tajima}, {Takahashi},
  {Tanaka}, {Thayer}, {Thayer}, {Thompson}, {Tibaldo}, {Tinivella}, {Tosti},
  {Troja}, {Usher}, {Vandenbroucke}, {Van Klaveren}, {Vasileiou}, {Vianello},
  {Vitale}, {Waite}, {Wallace}, {Winer}, {Wood}, {Wood}, {Wood}, {Yang}, \&
  {Zimmer}}]{LATPerformancePaper}
{Ackermann}, M., {Ajello}, M., {Albert}, A., {et~al.} 2012, \apjs, 203, 4

\bibitem[{{Ackermann} {et~al.}(2016){Ackermann}, {Ajello}, {Albert},
  {Anderson}, {Arimoto}, {Atwood}, {Axelsson}, {Baldini}, {Ballet},
  {Barbiellini}, {Baring}, {Bastieri}, {Becerra Gonzalez}, {Bellazzini},
  {Bissaldi}, {Blandford}, {Bloom}, {Bonino}, {Bottacini}, {Brandt}, {Bregeon},
  {Britto}, {Bruel}, {Buehler}, {Burnett}, {Buson}, {Caliandro}, {Cameron},
  {Caputo}, {Caragiulo}, {Caraveo}, {Casandjian}, {Cavazzuti}, {Charles},
  {Chekhtman}, {Chiang}, {Chiaro}, {Ciprini}, {Cohen-Tanugi}, {Cominsky},
  {Condon}, {Costanza}, {Cuoco}, {Cutini}, {D'Ammando}, {de Palma}, {Desiante},
  {Digel}, {Di Lalla}, {Di Mauro}, {Di Venere}, {Dom{\'{\i}}nguez}, {Drell},
  {Dubois}, {Dumora}, {Favuzzi}, {Fegan}, {Ferrara}, {Franckowiak}, {Fukazawa},
  {Funk}, {Fusco}, {Gargano}, {Gasparrini}, {Gehrels}, {Giglietto}, {Giomi},
  {Giommi}, {Giordano}, {Giroletti}, {Glanzman}, {Godfrey}, {Gomez-Vargas},
  {Granot}, {Green}, {Grenier}, {Grondin}, {Grove}, {Guillemot}, {Guiriec},
  {Hadasch}, {Harding}, {Hays}, {Hewitt}, {Hill}, {Horan}, {Jogler},
  {J{\'o}hannesson}, {Kamae}, {Kensei}, {Kocevski}, {Kuss}, {La Mura},
  {Larsson}, {Latronico}, {Lemoine-Goumard}, {Li}, {Li}, {Longo}, {Loparco},
  {Lovellette}, {Lubrano}, {Madejski}, {Magill}, {Maldera}, {Manfreda},
  {Marelli}, {Mayer}, {Mazziotta}, {McEnery}, {Meyer}, {Michelson}, {Mirabal},
  {Mizuno}, {Moiseev}, {Monzani}, {Moretti}, {Morselli}, {Moskalenko},
  {Murgia}, {Negro}, {Nuss}, {Ohsugi}, {Omodei}, {Orienti}, {Orlando}, {Ormes},
  {Paneque}, {Perkins}, {Pesce-Rollins}, {Piron}, {Pivato}, {Porter},
  {Racusin}, {Rain{\`o}}, {Rando}, {Razzaque}, {Reimer}, {Reimer}, {Reposeur},
  {Ritz}, {Rochester}, {Romani}, {Saz Parkinson}, {Sgr{\`o}}, {Simone},
  {Siskind}, {Smith}, {Spada}, {Spandre}, {Spinelli}, {Suson}, {Tajima},
  {Thayer}, {Thayer}, {Thompson}, {Tibaldo}, {Torres}, {Troja}, {Uchiyama},
  {Venters}, {Vianello}, {Wood}, {Wood}, {Zaharijas}, {Zhu}, \&
  {Zimmer}}]{Ackermann2016}
---. 2016, \apjl, 823, L2

\bibitem[{{Arnaud}(1996)}]{Arnaud1996}
{Arnaud}, K.~A. 1996, in Astronomical Society of the Pacific Conference Series,
  Vol. 101, Astronomical Data Analysis Software and Systems V, ed. G.~H.
  {Jacoby} \& J.~{Barnes}, 17

\bibitem[{{Atwood} {et~al.}(2009){Atwood}, {Abdo}, {Ackermann}, {Althouse},
  {Anderson}, {Axelsson}, {Baldini}, {Ballet}, {Band}, {Barbiellini}, \&
  et~al.}]{Atwood2009}
{Atwood}, W.~B., {Abdo}, A.~A., {Ackermann}, M., {et~al.} 2009, \apj, 697, 1071

\bibitem[{{Band} {et~al.}(1993){Band}, {Matteson}, {Ford}, {Schaefer},
  {Palmer}, {Teegarden}, {Cline}, {Briggs}, {Paciesas}, {Pendleton}, {Fishman},
  {Kouveliotou}, {Meegan}, {Wilson}, \& {Lestrade}}]{Band1993}
{Band}, D., {Matteson}, J., {Ford}, L., {et~al.} 1993, \apj, 413, 281

\bibitem[{{Barthelmy} {et~al.}(2005){Barthelmy}, {Barbier}, {Cummings},
  {Fenimore}, {Gehrels}, {Hullinger}, {Krimm}, {Markwardt}, {Palmer},
  {Parsons}, {Sato}, {Suzuki}, {Takahashi}, {Tashiro}, \&
  {Tueller}}]{Barthelmy2005}
{Barthelmy}, S.~D., {Barbier}, L.~M., {Cummings}, J.~R., {et~al.} 2005, \ssr,
  120, 143

\bibitem[{{Berger}(2014)}]{Berger2014}
{Berger}, E. 2014, \araa, 52, 43

\bibitem[{{Bhattacharyya}(2017)}]{Bhatt2017}
{Bhattacharyya}, S. 2017, arXiv e-prints, arXiv:1711.09083

\bibitem[{Blackburn {et~al.}(2015)Blackburn, Briggs, Camp, Christensen,
  Connaughton, Jenke, Remillard, \& Veitch}]{blackburn2015high}
Blackburn, L., Briggs, M.~S., Camp, J., {et~al.} 2015, The Astrophysical
  Journal Supplement Series, 217, 8

\bibitem[{{Bregman}(2007)}]{Bregman2007}
{Bregman}, J.~N. 2007, \araa, 45, 221

\bibitem[{{Buehler} {et~al.}(2012){Buehler}, {Scargle}, {Blandford}, {Baldini},
  {Baring}, {Belfiore}, {Charles}, {Chiang}, {D'Ammando}, {Dermer}, {Funk},
  {Grove}, {Harding}, {Hays}, {Kerr}, {Massaro}, {Mazziotta}, {Romani}, {Saz
  Parkinson}, {Tennant}, \& {Weisskopf}}]{Buehler2012}
{Buehler}, R., {Scargle}, J.~D., {Blandford}, R.~D., {et~al.} 2012, \apj, 749,
  26

\bibitem[{{B{\"u}hler} \& {Blandford}(2014)}]{Buhler2014}
{B{\"u}hler}, R., \& {Blandford}, R. 2014, Reports on Progress in Physics, 77,
  066901

\bibitem[{{Callister} {et~al.}(2016){Callister}, {Kanner}, \&
  {Weinstein}}]{Callister2016}
{Callister}, T., {Kanner}, J., \& {Weinstein}, A. 2016, \apj, 825, L12

\bibitem[{{Champion} {et~al.}(2016){Champion}, {Petroff}, {Kramer}, {Keith},
  {Bailes}, {Barr}, {Bates}, {Bhat}, {Burgay}, {Burke-Spolaor}, {Flynn},
  {Jameson}, {Johnston}, {Ng}, {Levin}, {Possenti}, {Stappers}, {van Straten},
  {Thornton}, {Tiburzi}, \& {Lyne}}]{Champion2016}
{Champion}, D.~J., {Petroff}, E., {Kramer}, M., {et~al.} 2016, \mnras, 460, L30

\bibitem[{{Chatterjee} {et~al.}(2017){Chatterjee}, {Law}, {Wharton},
  {Burke-Spolaor}, {Hessels}, {Bower}, {Cordes}, {Tendulkar}, {Bassa},
  {Demorest}, {Butler}, {Seymour}, {Scholz}, {Abruzzo}, {Bogdanov}, {Kaspi},
  {Keimpema}, {Lazio}, {Marcote}, {McLaughlin}, {Paragi}, {Ransom}, {Rupen},
  {Spitler}, \& {van Langevelde}}]{Chatterjee2017}
{Chatterjee}, S., {Law}, C.~J., {Wharton}, R.~S., {et~al.} 2017, \nat, 541, 58

\bibitem[{{Connaughton} {et~al.}(2015){Connaughton}, {Briggs}, {Goldstein},
  {Meegan}, {Paciesas}, {Preece}, {Wilson-Hodge}, {Gibby}, {Greiner}, {Gruber},
  {Jenke}, {Kippen}, {Pelassa}, {Xiong}, {Yu}, {Bhat}, {Burgess}, {Byrne},
  {Fitzpatrick}, {Foley}, {Giles}, {Guiriec}, {van der Horst}, {von Kienlin},
  {McBreen}, {McGlynn}, {Tierney}, \& {Zhang}}]{Connaughton2015}
{Connaughton}, V., {Briggs}, M.~S., {Goldstein}, A., {et~al.} 2015, \apjs, 216,
  32

\bibitem[{{Cordes} \& {Lazio}(2002)}]{CordesLazio2002}
{Cordes}, J.~M., \& {Lazio}, T.~J.~W. 2002, ArXiv Astrophysics e-prints,
  astro-ph/0207156

\bibitem[{{Cordes} \& {Wasserman}(2016)}]{CordesWasserman2016}
{Cordes}, J.~M., \& {Wasserman}, I. 2016, \mnras, 457, 232

\bibitem[{{DeLaunay} {et~al.}(2016){DeLaunay}, {Fox}, {Murase},
  {M{\'e}sz{\'a}ros}, {Keivani}, {Messick}, {Mostaf{\'a}}, {Oikonomou}, {Te{\v
  s}i{\'c}}, \& {Turley}}]{DeLaunay2016}
{DeLaunay}, J.~J., {Fox}, D.~B., {Murase}, K., {et~al.} 2016, \apjl, 832, L1

\bibitem[{{Dokuchaev} \& {Eroshenko}(2017)}]{Dokuchaev2017}
{Dokuchaev}, V.~I., \& {Eroshenko}, Y.~N. 2017, ArXiv e-prints,
  arXiv:1701.02492

\bibitem[{{Duncan} \& {Thompson}(1992)}]{Duncan1992}
{Duncan}, R.~C., \& {Thompson}, C. 1992, \apj, 392, L9

\bibitem[{{Falcke} \& {Rezzolla}(2014)}]{Falke2014}
{Falcke}, H., \& {Rezzolla}, L. 2014, \aap, 562, A137

\bibitem[{{Gehrels} {et~al.}(2004){Gehrels}, {Chincarini}, {Giommi}, {Mason},
  {Nousek}, {Wells}, {White}, {Barthelmy}, {Burrows}, {Cominsky}, {Hurley},
  {Marshall}, {M{\'e}sz{\'a}ros}, {Roming}, {Angelini}, {Barbier}, {Belloni},
  {Campana}, {Caraveo}, {Chester}, {Citterio}, {Cline}, {Cropper}, {Cummings},
  {Dean}, {Feigelson}, {Fenimore}, {Frail}, {Fruchter}, {Garmire}, {Gendreau},
  {Ghisellini}, {Greiner}, {Hill}, {Hunsberger}, {Krimm}, {Kulkarni}, {Kumar},
  {Lebrun}, {Lloyd- Ronning}, {Markwardt}, {Mattson}, {Mushotzky}, {Norris},
  {Osborne}, {Paczynski}, {Palmer}, {Park}, {Parsons}, {Paul}, {Rees},
  {Reynolds}, {Rhoads}, {Sasseen}, {Schaefer}, {Short}, {Smale}, {Smith},
  {Stella}, {Tagliaferri}, {Takahashi}, {Tashiro}, {Townsley}, {Tueller},
  {Turner}, {Vietri}, {Voges}, {Ward}, {Willingale}, {Zerbi}, \&
  {Zhang}}]{Gehrels2004}
{Gehrels}, N., {Chincarini}, G., {Giommi}, P., {et~al.} 2004, \apj, 611, 1005

\bibitem[{{Goldstein} {et~al.}(2016){Goldstein}, {Burns}, {Hamburg},
  {Connaughton}, {Veres}, {Briggs}, {Hui}, \& {The GBM-LIGO
  Collaboration}}]{2016arXiv161202395G}
{Goldstein}, A., {Burns}, E., {Hamburg}, R., {et~al.} 2016, ArXiv e-prints,
  arXiv:1612.02395

\bibitem[{{Gu} {et~al.}(2016){Gu}, {Dong}, {Liu}, {Ma}, \& {Wang}}]{Gu2016}
{Gu}, W.-M., {Dong}, Y.-Z., {Liu}, T., {Ma}, R., \& {Wang}, J. 2016, \apj, 823,
  L28

\bibitem[{{Hankins} \& {Eilek}(2007)}]{Hankins2007}
{Hankins}, T.~H., \& {Eilek}, J.~A. 2007, \apj, 670, 693

\bibitem[{{Hurley} {et~al.}(1999){Hurley}, {Cline}, {Mazets}, {Barthelmy},
  {Butterworth}, {Marshall}, {Palmer}, {Aptekar}, {Golenetskii}, {Il'Inskii},
  {Frederiks}, {McTiernan}, {Gold}, \& {Trombka}}]{Hurley1999}
{Hurley}, K., {Cline}, T., {Mazets}, E., {et~al.} 1999, \nat, 397, 41

\bibitem[{{Hurley} {et~al.}(2005){Hurley}, {Boggs}, {Smith}, {Duncan}, {Lin},
  {Zoglauer}, {Krucker}, {Hurford}, {Hudson}, {Wigger}, {Hajdas}, {Thompson},
  {Mitrofanov}, {Sanin}, {Boynton}, {Fellows}, {von Kienlin}, {Lichti}, {Rau},
  \& {Cline}}]{Hurley2005}
{Hurley}, K., {Boggs}, S.~E., {Smith}, D.~M., {et~al.} 2005, \nat, 434, 1098

\bibitem[{{Katz}(2018)}]{Katz2018}
{Katz}, J.~I. 2018, Progress in Particle and Nuclear Physics, 103, 1

\bibitem[{{Keane} \& {Petroff}(2015)}]{KeanePetroff2015}
{Keane}, E.~F., \& {Petroff}, E. 2015, \mnras, 447, 2852

\bibitem[{{Keane} {et~al.}(2016){Keane}, {Johnston}, {Bhandari}, {Barr},
  {Bhat}, {Burgay}, {Caleb}, {Flynn}, {Jameson}, {Kramer}, {Petroff},
  {Possenti}, {van Straten}, {Bailes}, {Burke-Spolaor}, {Eatough}, {Stappers},
  {Totani}, {Honma}, {Furusawa}, {Hattori}, {Morokuma}, {Niino}, {Sugai},
  {Terai}, {Tominaga}, {Yamasaki}, {Yasuda}, {Allen}, {Cooke}, {Jencson},
  {Kasliwal}, {Kaplan}, {Tingay}, {Williams}, {Wayth}, {Chandra}, {Perrodin},
  {Berezina}, {Mickaliger}, \& {Bassa}}]{Keane2016}
{Keane}, E.~F., {Johnston}, S., {Bhandari}, S., {et~al.} 2016, \nat, 530, 453

\bibitem[{{Kouveliotou} {et~al.}(1993){Kouveliotou}, {Meegan}, {Fishman},
  {Bhat}, {Briggs}, {Koshut}, {Paciesas}, \& {Pendleton}}]{Kouveliotou1993}
{Kouveliotou}, C., {Meegan}, C.~A., {Fishman}, G.~J., {et~al.} 1993, \apjl,
  413, L101

\bibitem[{{Kouveliotou} {et~al.}(1998){Kouveliotou}, {Dieters}, {Strohmayer},
  {van Paradijs}, {Fishman}, {Meegan}, {Hurley}, {Kommers}, {Smith}, {Frail},
  \& {Murakami}}]{Kouveliotou1998}
{Kouveliotou}, C., {Dieters}, S., {Strohmayer}, T., {et~al.} 1998, \nat, 393,
  235

\bibitem[{{Krimm} {et~al.}(2013){Krimm}, {Holland}, {Corbet}, {Pearlman},
  {Romano}, {Kennea}, {Bloom}, {Barthelmy}, {Baumgartner}, {Cummings},
  {Gehrels}, {Lien}, {Markwardt}, {Palmer}, {Sakamoto}, {Stamatikos}, \&
  {Ukwatta}}]{Krimm2013}
{Krimm}, H.~A., {Holland}, S.~T., {Corbet}, R.~H.~D., {et~al.} 2013, \apjs,
  209, 14

\bibitem[{{Kulkarni} {et~al.}(2015){Kulkarni}, {Ofek}, \&
  {Neill}}]{Kulkarni2015}
{Kulkarni}, S.~R., {Ofek}, E.~O., \& {Neill}, J.~D. 2015, ArXiv e-prints,
  arXiv:1511.09137

\bibitem[{{Li} {et~al.}(2018){Li}, {Huang}, {Geng}, \& {Li}}]{Li2018}
{Li}, L.-B., {Huang}, Y.-F., {Geng}, J.-J., \& {Li}, B. 2018, Research in
  Astronomy and Astrophysics, 18, 061

\bibitem[{{Li} {et~al.}(2011){Li}, {Chornock}, {Leaman}, {Filippenko},
  {Poznanski}, {Wang}, {Ganeshalingam}, \& {Mannucci}}]{Li2011}
{Li}, W., {Chornock}, R., {Leaman}, J., {et~al.} 2011, \mnras, 412, 1473

\bibitem[{{Lien} {et~al.}(2014){Lien}, {Sakamoto}, {Gehrels}, {Palmer},
  {Barthelmy}, {Graziani}, \& {Cannizzo}}]{Lien2014}
{Lien}, A., {Sakamoto}, T., {Gehrels}, N., {et~al.} 2014, \apj, 783, 24

\bibitem[{{Lin} {et~al.}(2018){Lin}, {Cheng}, \& {Gan}}]{Lin2018}
{Lin}, Y., {Cheng}, Z., \& {Gan}, L. 2018, Scientia Sinica Physica, Mechanica
  \&amp; Astronomica, 48, 029501

\bibitem[{{Lorimer}(2018)}]{Lorimer2018}
{Lorimer}, D.~R. 2018, Nature Astronomy, 2, 860

\bibitem[{{Lorimer} {et~al.}(2007){Lorimer}, {Bailes}, {McLaughlin},
  {Narkevic}, \& {Crawford}}]{Lorimer2007}
{Lorimer}, D.~R., {Bailes}, M., {McLaughlin}, M.~A., {Narkevic}, D.~J., \&
  {Crawford}, F. 2007, Science, 318, 777

\bibitem[{{Lyubarsky}(2014)}]{Lyubarsky2014}
{Lyubarsky}, Y. 2014, \mnras, 442, L9

\bibitem[{{Lyutikov}(2002)}]{Lyutikov2002}
{Lyutikov}, M. 2002, \apjl, 580, L65

\bibitem[{{Manchester} {et~al.}(2016){Manchester}, {Hobbs}, {Teoh}, \&
  {Hobbs}}]{Manchester2016}
{Manchester}, R.~N., {Hobbs}, G.~B., {Teoh}, A., \& {Hobbs}, M. 2016, VizieR
  Online Data Catalog, B/psr

\bibitem[{{Maoz} {et~al.}(2015){Maoz}, {Loeb}, {Shvartzvald}, {Sitek}, {Engel},
  {Kiefer}, {Kiraga}, {Levi}, {Mazeh}, {Pawlak}, {Rich}, {Tal-Or}, \&
  {Wyrzykowski}}]{Maoz2015}
{Maoz}, D., {Loeb}, A., {Shvartzvald}, Y., {et~al.} 2015, \mnras, 454, 2183

\bibitem[{{Marcote} {et~al.}(2017){Marcote}, {Paragi}, {Hessels}, {Keimpema},
  {van Langevelde}, {Huang}, {Bassa}, {Bogdanov}, {Bower}, {Burke-Spolaor},
  {Butler}, {Campbell}, {Chatterjee}, {Cordes}, {Demorest}, {Garrett}, {Ghosh},
  {Kaspi}, {Law}, {Lazio}, {McLaughlin}, {Ransom}, {Salter}, {Scholz},
  {Seymour}, {Siemion}, {Spitler}, {Tendulkar}, \& {Wharton}}]{Marcote2017}
{Marcote}, B., {Paragi}, Z., {Hessels}, J.~W.~T., {et~al.} 2017, \apjl, 834, L8

\bibitem[{{Mazets} {et~al.}(1979){Mazets}, {Golentskii}, {Ilinskii}, {Aptekar},
  \& {Guryan}}]{Mazets1979}
{Mazets}, E.~P., {Golentskii}, S.~V., {Ilinskii}, V.~N., {Aptekar}, R.~L., \&
  {Guryan}, I.~A. 1979, \nat, 282, 587

\bibitem[{{McQuinn}(2014)}]{McQuinn2014}
{McQuinn}, M. 2014, \apjl, 780, L33

\bibitem[{{Meegan} {et~al.}(2009){Meegan}, {Lichti}, {Bhat}, {Bissaldi},
  {Briggs}, {Connaughton}, {Diehl}, {Fishman}, {Greiner}, {Hoover}, {van der
  Horst}, {von Kienlin}, {Kippen}, {Kouveliotou}, {McBreen}, {Paciesas},
  {Preece}, {Steinle}, {Wallace}, {Wilson}, \& {Wilson-Hodge}}]{Meegan2009}
{Meegan}, C., {Lichti}, G., {Bhat}, P.~N., {et~al.} 2009, \apj, 702, 791

\bibitem[{{Metzger} {et~al.}(2017){Metzger}, {Berger}, \&
  {Margalit}}]{Metzger2017}
{Metzger}, B.~D., {Berger}, E., \& {Margalit}, B. 2017, \apj, 841, 14

\bibitem[{{Mickaliger} {et~al.}(2012){Mickaliger}, {McLaughlin}, {Lorimer},
  {Langston}, {Bilous}, {Kondratiev}, {Lyutikov}, {Ransom}, \&
  {Palliyaguru}}]{Mickaliger2012}
{Mickaliger}, M.~B., {McLaughlin}, M.~A., {Lorimer}, D.~R., {et~al.} 2012,
  \apj, 760, 64

\bibitem[{{Nicholl} {et~al.}(2017){Nicholl}, {Williams}, {Berger}, {Villar},
  {Alexander}, {Eftekhari}, \& {Metzger}}]{Nicholl2017}
{Nicholl}, M., {Williams}, P.~K.~G., {Berger}, E., {et~al.} 2017, \apj, 843, 84

\bibitem[{{Palaniswamy} {et~al.}(2018){Palaniswamy}, {Li}, \&
  {Zhang}}]{Palaniswamy2018}
{Palaniswamy}, D., {Li}, Y., \& {Zhang}, B. 2018, \apj, 854, L12

\bibitem[{{Palmer} {et~al.}(2005){Palmer}, {Barthelmy}, {Gehrels}, {Kippen},
  {Cayton}, {Kouveliotou}, {Eichler}, {Wijers}, {Woods}, {Granot}, {Lyubarsky},
  {Ramirez-Ruiz}, {Barbier}, {Chester}, {Cummings}, {Fenimore}, {Finger},
  {Gaensler}, {Hullinger}, {Krimm}, {Markwardt}, {Nousek}, {Parsons}, {Patel},
  {Sakamoto}, {Sato}, {Suzuki}, \& {Tueller}}]{Palmer2005}
{Palmer}, D.~M., {Barthelmy}, S., {Gehrels}, N., {et~al.} 2005, \nat, 434, 1107

\bibitem[{{Pen} \& {Connor}(2015)}]{Pen2015}
{Pen}, U.-L., \& {Connor}, L. 2015, \apj, 807, 179

\bibitem[{{Petroff} {et~al.}(2016){Petroff}, {Barr}, {Jameson}, {Keane},
  {Bailes}, {Kramer}, {Morello}, {Tabbara}, \& {van Straten}}]{Petroff2016}
{Petroff}, E., {Barr}, E.~D., {Jameson}, A., {et~al.} 2016, Publications of the
  Astronomical Society of Australia, 33, e045

\bibitem[{{Popov} \& {Postnov}(2013)}]{Popov2013}
{Popov}, S.~B., \& {Postnov}, K.~A. 2013, ArXiv e-prints, arXiv:1307.4924

\bibitem[{{Readhead}(1994)}]{Readhead1994}
{Readhead}, A. C.~S. 1994, \apj, 426, 51

\bibitem[{{Savchenko} {et~al.}(2018{\natexlab{a}}){Savchenko}, {Ferrigno},
  {Panessa}, {Bazzano}, {Ubertini}, {Kuulkers}, \& {Keane}}]{2018ATel11431}
{Savchenko}, V., {Ferrigno}, C., {Panessa}, F., {et~al.} 2018{\natexlab{a}},
  The Astronomer's Telegram, 11431, 1

\bibitem[{{Savchenko} {et~al.}(2018{\natexlab{b}}){Savchenko}, {Ferrigno},
  {Panessa}, {Bazzano}, {Ubertini}, \& {Keane}}]{2018ATel11387}
---. 2018{\natexlab{b}}, The Astronomer's Telegram, 11387, 1

\bibitem[{{Savchenko} {et~al.}(2018{\natexlab{c}}){Savchenko}, {Panessa},
  {Ferrigno}, {Keane}, {Bazzano}, {Burgay}, {Kuulkers}, {Petroff}, {Ubertini},
  \& {Diehl}}]{2018ATel11386}
{Savchenko}, V., {Panessa}, F., {Ferrigno}, C., {et~al.} 2018{\natexlab{c}},
  The Astronomer's Telegram, 11386, 1

\bibitem[{{Scholz} {et~al.}(2016){Scholz}, {Spitler}, {Hessels}, {Chatterjee},
  {Cordes}, {Kaspi}, {Wharton}, {Bassa}, {Bogdanov}, {Camilo}, {Crawford},
  {Deneva}, {van Leeuwen}, {Lynch}, {Madsen}, {McLaughlin}, {Mickaliger},
  {Parent}, {Patel}, {Ransom}, {Seymour}, {Stairs}, {Stappers}, \&
  {Tendulkar}}]{Scholz2016}
{Scholz}, P., {Spitler}, L.~G., {Hessels}, J.~W.~T., {et~al.} 2016, \apj, 833,
  177

\bibitem[{{Scholz} {et~al.}(2017{\natexlab{a}}){Scholz}, {Bogdanov}, {Hessels},
  {Lynch}, {Spitler}, {Bassa}, {Bower}, {Burke-Spolaor}, {Butler},
  {Chatterjee}, {Cordes}, {Gourdji}, {Kaspi}, {Law}, {Marcote}, {McLaughlin},
  {Michilli}, {Paragi}, {Ransom}, {Seymour}, {Tendulkar}, \&
  {Wharton}}]{Scholz2017}
{Scholz}, P., {Bogdanov}, S., {Hessels}, J.~W.~T., {et~al.} 2017{\natexlab{a}},
  \apj, 846, 80

\bibitem[{{Scholz} {et~al.}(2017{\natexlab{b}}){Scholz}, {Bogdanov}, {Hessels},
  {Lynch}, {Spitler}, {Bassa}, {Bower}, {Burke-Spolaor}, {Butler},
  {Chatterjee}, {Cordes}, {Gourdji}, {Kaspi}, {Law}, {Marcote}, {McLaughlin},
  {Michilli}, {Paragi}, {Ransom}, {Seymour}, {Tendulkar}, \&
  {Wharton}}]{Scholz2017b}
---. 2017{\natexlab{b}}, \apj, 846, doi:10.3847/1538-4357/aa8456

\bibitem[{{Shull} {et~al.}(2012){Shull}, {Smith}, \& {Danforth}}]{Shull2012}
{Shull}, J.~M., {Smith}, B.~D., \& {Danforth}, C.~W. 2012, \apj, 759, 23

\bibitem[{{Spitler} {et~al.}(2016){Spitler}, {Scholz}, {Hessels}, {Bogdanov},
  {Brazier}, {Camilo}, {Chatterjee}, {Cordes}, {Crawford}, {Deneva}, {Ferdman},
  {Freire}, {Kaspi}, {Lazarus}, {Lynch}, {Madsen}, {McLaughlin}, {Patel},
  {Ransom}, {Seymour}, {Stairs}, {Stappers}, {van Leeuwen}, \&
  {Zhu}}]{Spitler2016}
{Spitler}, L.~G., {Scholz}, P., {Hessels}, J.~W.~T., {et~al.} 2016, \nat, 531,
  202

\bibitem[{{Striani} {et~al.}(2011){Striani}, {Tavani}, {Piano}, {Donnarumma},
  {Pucella}, {Vittorini}, {Bulgarelli}, {Trois}, {Pittori}, {Verrecchia},
  {Costa}, {Weisskopf}, {Tennant}, {Argan}, {Barbiellini}, {Caraveo},
  {Cardillo}, {Cattaneo}, {Chen}, {De Paris}, {Del Monte}, {Di Cocco},
  {Evangelista}, {Ferrari}, {Feroci}, {Fuschino}, {Galli}, {Gianotti},
  {Giuliani}, {Labanti}, {Lapshov}, {Lazzarotto}, {Longo}, {Marisaldi},
  {Mereghetti}, {Morselli}, {Pacciani}, {Pellizzoni}, {Perotti}, {Picozza},
  {Pilia}, {Rapisarda}, {Rappoldi}, {Sabatini}, {Soffitta}, {Trifoglio},
  {Vercellone}, {Lucarelli}, {Santolamazza}, \& {Giommi}}]{Striani2011}
{Striani}, E., {Tavani}, M., {Piano}, G., {et~al.} 2011, \apjl, 741, L5

\bibitem[{{Takami} {et~al.}(2014){Takami}, {Kyutoku}, \& {Ioka}}]{Takami2014}
{Takami}, H., {Kyutoku}, K., \& {Ioka}, K. 2014, \prd, 89, 063006

\bibitem[{{Tendulkar} {et~al.}(2016){Tendulkar}, {Kaspi}, \&
  {Patel}}]{Tendulkar2016}
{Tendulkar}, S.~P., {Kaspi}, V.~M., \& {Patel}, C. 2016, \apj, 827, 59

\bibitem[{{Tendulkar} {et~al.}(2017){Tendulkar}, {Bassa}, {Cordes}, {Bower},
  {Law}, {Chatterjee}, {Adams}, {Bogdanov}, {Burke-Spolaor}, {Butler},
  {Demorest}, {Hessels}, {Kaspi}, {Lazio}, {Maddox}, {Marcote}, {McLaughlin},
  {Paragi}, {Ransom}, {Scholz}, {Seymour}, {Spitler}, {van Langevelde}, \&
  {Wharton}}]{Tendulkar2017}
{Tendulkar}, S.~P., {Bassa}, C.~G., {Cordes}, J.~M., {et~al.} 2017, \apjl, 834,
  L7

\bibitem[{{The CHIME/FRB Collaboration} {et~al.}(2019){The CHIME/FRB
  Collaboration}, {:}, {Amiri}, {Bandura}, {Bhardwaj}, {Boubel}, {Boyce},
  {Boyle}, {Brar}, {Burhanpurkar}, {Cassanelli}, {Chawla}, {Cliche},
  {Cubranic}, {Deng}, {Denman}, {Dobbs}, {Fandino}, {Fonseca}, {Gaensler},
  {Gilbert}, {Gill}, {Giri}, {Good}, {Halpern}, {Hanna}, {Hill}, {Hinshaw},
  {H{\"o}fer}, {Josephy}, {Kaspi}, {Landecker}, {Lang}, {Lin}, {Masui},
  {Mckinven}, {Mena-Parra}, {Merryfield}, {Michilli}, {Milutinovic}, {Moatti},
  {Naidu}, {Newburgh}, {Ng}, {Patel}, {Pen}, {Pinsonneault-Marotte}, {Pleunis},
  {Rafiei-Ravandi}, {Rahman}, {Ransom}, {Renard}, {Scholz}, {Shaw}, {Siegel},
  {Smith}, {Stairs}, {Tendulkar}, {Tretyakov}, {Vanderlinde}, \&
  {Yadav}}]{Amiri2019}
{The CHIME/FRB Collaboration}, {:}, {Amiri}, M., {et~al.} 2019, arXiv e-prints,
  arXiv:1901.04525

\bibitem[{{Thornton} {et~al.}(2013){Thornton}, {Stappers}, {Bailes},
  {Barsdell}, {Bates}, {Bhat}, {Burgay}, {Burke-Spolaor}, {Champion}, {Coster},
  {D'Amico}, {Jameson}, {Johnston}, {Keith}, {Kramer}, {Levin}, {Milia}, {Ng},
  {Possenti}, \& {van Straten}}]{Thornton2013}
{Thornton}, D., {Stappers}, B., {Bailes}, M., {et~al.} 2013, Science, 341, 53

\bibitem[{{Vieyro} {et~al.}(2017){Vieyro}, {Romero}, {Bosch-Ramon}, {Marcote},
  \& {del Valle}}]{Vieyro2017}
{Vieyro}, F.~L., {Romero}, G.~E., {Bosch-Ramon}, V., {Marcote}, B., \& {del
  Valle}, M.~V. 2017, \aap, 602, A64

\bibitem[{{Williams} \& {Berger}(2016)}]{Williams2016}
{Williams}, P.~K.~G., \& {Berger}, E. 2016, \apjl, 821, L22

\bibitem[{{Zhang}(2014)}]{Zhang2014}
{Zhang}, B. 2014, \apj, 780, L21

\bibitem[{{Zhang}(2016)}]{Zhang2016}
---. 2016, \apj, 827, L31

\bibitem[{{Zhang}(2017)}]{Zhang2017}
---. 2017, \apjl, 836, L32

\bibitem[{{Zhou} {et~al.}(2014){Zhou}, {Li}, {Wang}, {Fan}, \&
  {Wei}}]{Zhou2014}
{Zhou}, B., {Li}, X., {Wang}, T., {Fan}, Y.-Z., \& {Wei}, D.-M. 2014, \prd, 89,
  107303

\end{thebibliography}

\appendix
\section{A.1 FRB Observations by Instrument}
Table \ref{tab:FRBs} summarizes which FRBs were observable within the FOV of the \textit{Fermi} GBM, \textit{Fermi} LAT, and \textit{Swift} BAT at the time of radio detection. Here Y denotes the FRB was observed by the instrument and N denotes it was not observed. The subsequent repeating bursts from FRB121102 have been combined for brevity. FRB131104 was detected on the edge of the BAT FOV but is excluded due to its low partial coding fraction.  \\
\clearpage
\begin{table*}[h]
\centering
\caption{Summary of observations available per FRB.}
\begin{tabular}{cccc}
 FRB Name & \textit{Fermi} GBM & \textit{Fermi} LAT & \textit{Swift} BAT    \\ \hline \hline
010125& N & N & N  \\
010621& N & N & N  \\
010724& N & N & N  \\
090625& Y & Y & N  \\
110220& N & N & N  \\
110523& Y & N & N  \\
110626& Y & N & N  \\
110703& Y & N & N  \\
120127& N & N & N  \\
121002& N & N & N  \\
121102& Y & Y & N  \\
130626& N & N & N  \\
130628& Y & Y & N  \\
130729& Y & N & N  \\
131104& Y & N & N  \\
140514& N & N & N  \\
150215& Y & Y & Y  \\
150418& Y & Y & N  \\
150807& Y & N & N  \\
160317& Y & Y & N  \\
160410& N & N & Y  \\
160608& Y & Y & N  \\
170107& N & N & N  \\ \hline


\end{tabular}
\label{tab:FRBs}
\end{table*}

\section{A.2 Summary of extraneous signals detected with the \textit{Fermi} GBM targeted search}
We follow a similar approach in our \textit{Fermi} GBM analysis to that of electromagnetic followup of gravitational wave compact binary sources. \citet{blackburn2015high} develop a Bayesian method to search GBM continuous data for coincident signals around LIGO triggers. The analysis only requires an event time for the LIGO trigger and a localization probability map. Therefore it is convenient to adapt the method to our purposes for candidates temporally coincident with FRBs. A log likelihood ratio (logLR) parameter is calculated for each FRB in the \textit{Fermi} GBM data to determine the probability of the presence of a signal compared to the null hypothesis of a constant background (see \citet{2016arXiv161202395G} for more details about this analysis). LogLR values greater than 10.0 are likely indicative of a real signal. We find six candidate signals coincident in time with the FRB detections ($t_{FRB}$) however we determine them all to be unrelated for the reasons outlined below:

\subsection{Local Particle Activity}
\textit{Fermi} is sensitive to increased levels of local particle activity in the magnetosphere along its orbit (even when outside the South Atlantic Anomaly).  These events are characterized by long (several tens to hundreds of seconds), smooth, and hard signals observed as a slow rise against the normal background emission. Three of the candidate signals (FRB131104 and bursts 19 and 20 from FRB121102) discovered by the targeted search are likely caused by this local magnetospheric activity. All three were identified within the 100 second timescale searches with logLR values greater than 10.0. None of the signals are temporally coincident with the FRBs. The signal around burst 19 occurred 90 seconds before $t_{FRB}$, the signal around burst 20 occurred 450 seconds before $t_{FRB}$, and third signal occurred  331 seconds after the $t_{FRB}$ of FRB131104. We considered all three candidate signals to be unrelated. \\

\subsection{Other Unrelated Signals}

\begin{figure}[h] 
     \begin{center}
        \subfigure[]{%
            \includegraphics[width=0.65\textwidth]{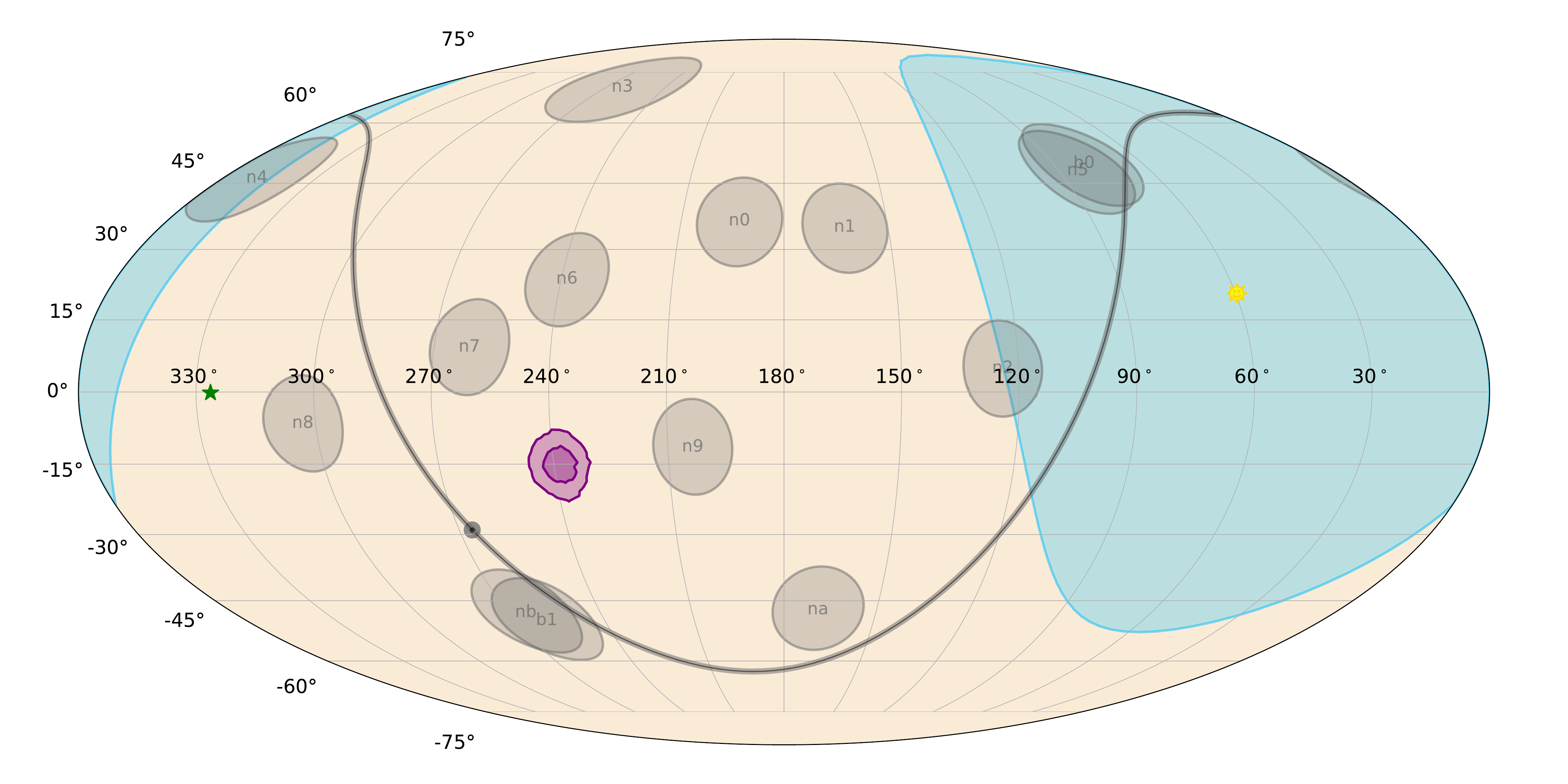} \label{fig:skymaps_a}
        }\\
        \subfigure[]{%
            \includegraphics[width=0.65\textwidth]{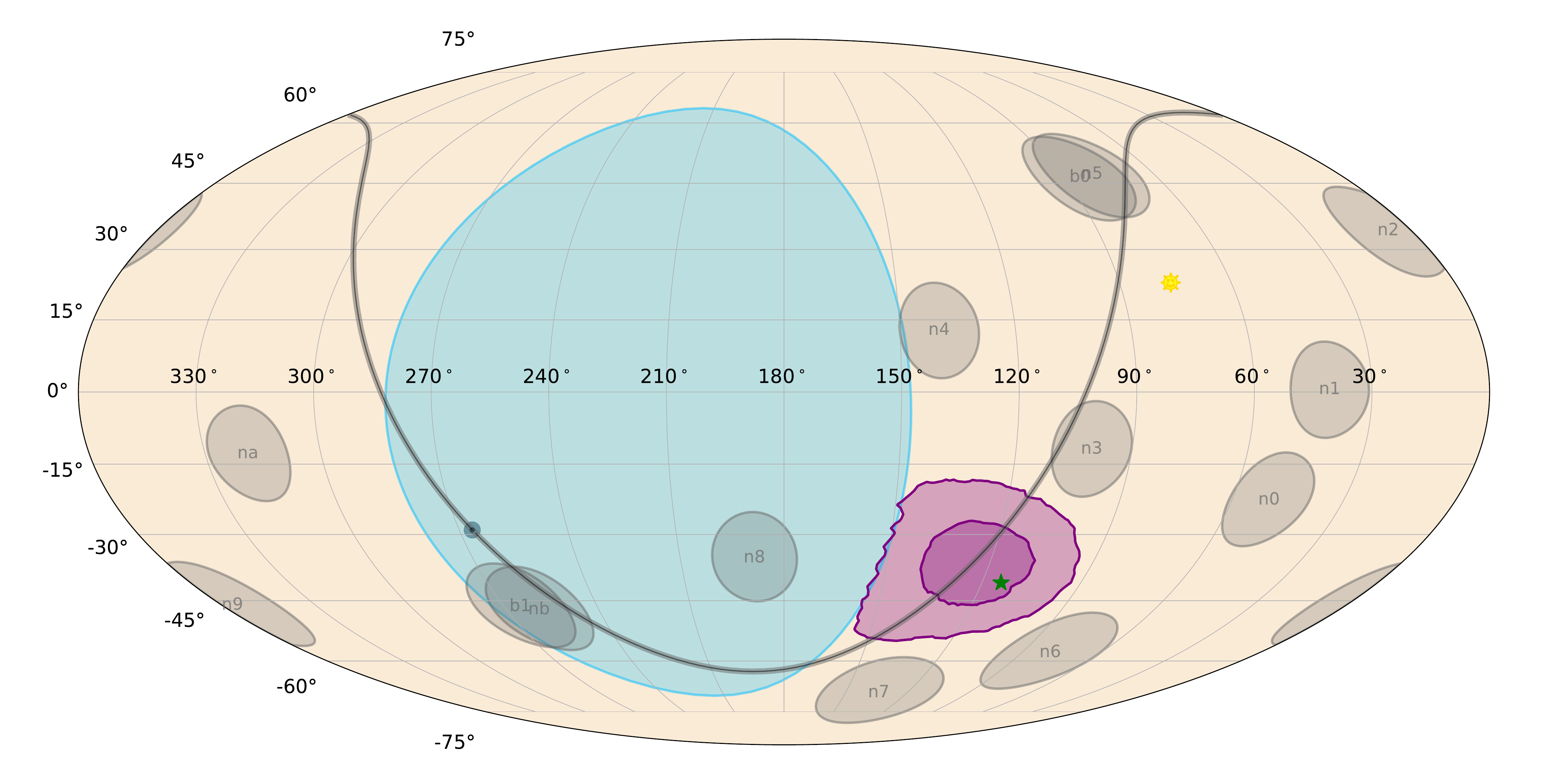}
            \label{fig:skymaps_b}
        }\\
        \subfigure[]{%
            \includegraphics[width=0.65\textwidth]{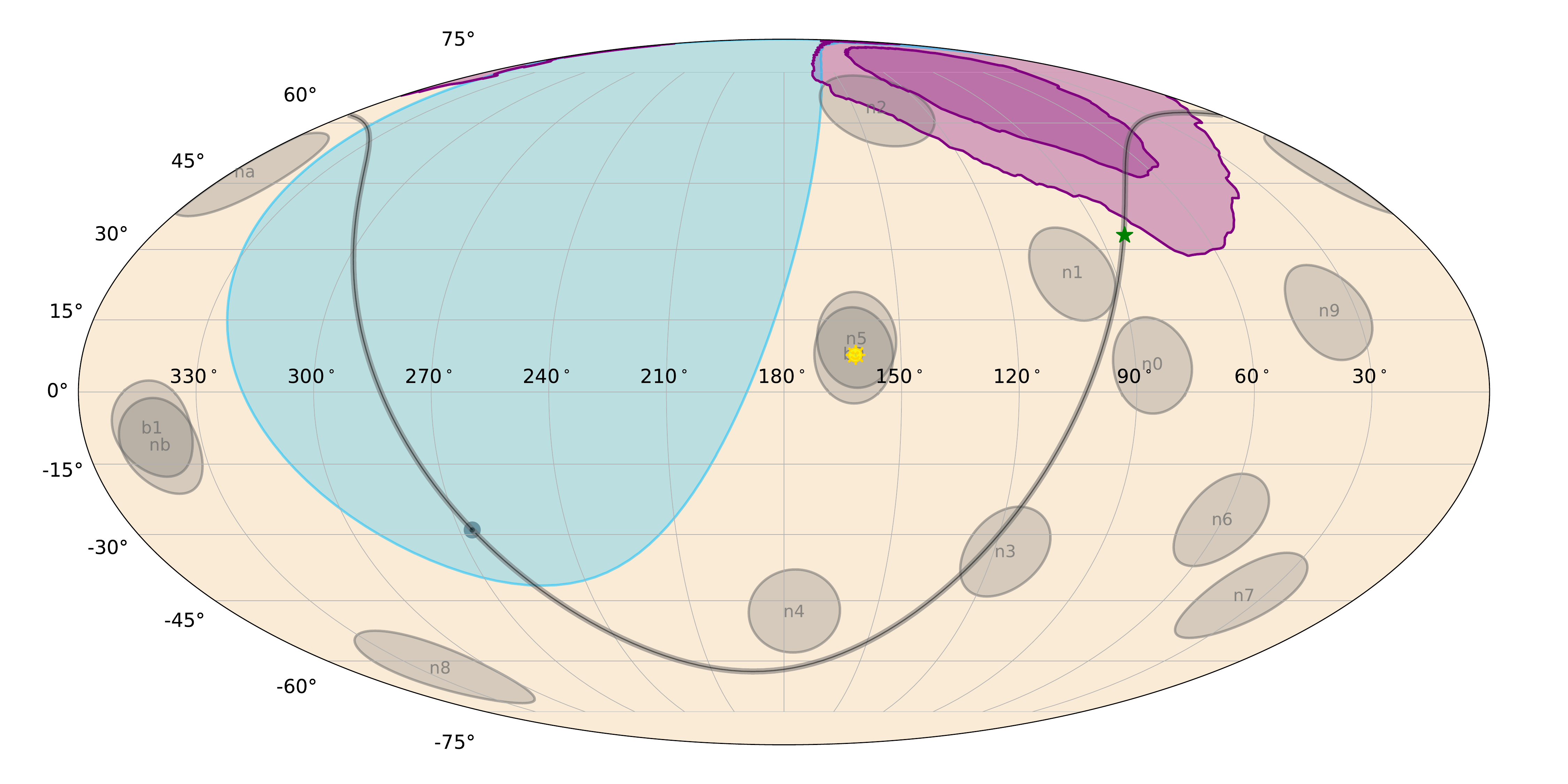}
            \label{fig:skymaps_c}
        }%
    \end{center}
    \caption{A localization map of the candidate signal for a.) FRB110523, b.) FRB160608, and c.) FRB121102 Burst 19. The blue shaded region represents the area occulted by the Earth, the gray stripe is the Galactic plane, the gray circles represent the positions of each of the detectors, and the yellow and green stars are the locations of the Sun and FRB, respectively. The purple region shows the 50\% and 90\% confidence regions for the GBM localization of the candidate signal.}
    \label{fig:skymaps}
\end{figure}

FRB110523: Only CTIME data are available for the analysis of FRB110523. A candidate signal is seen $\sim80$ seconds before $t_{FRB}$ with a duration of 8 seconds and a logLR of 22. The signal is soft and localizes near, but not on, the Galactic plane (Figure \ref{fig:skymaps_a}). The signal is not localized near the FRB.  We assume this to be unrelated particle activity. \\

FRB160608: A candidate signal is seen $\sim$10 seconds before $t_{FRB}$ with a logLR of 10.8. The signal properties are consistent with that of a Galactic transient (i.e., a soft, 10 second long burst that localizes to the Galactic plane). Although the localization is consistent with the FRB (Figure \ref{fig:skymaps_b}) based on further analysis with the BAT we suspect it is likely a flare from the nearby high-mass X-ray binary system Vela X-1.\\

FRB121102 Burst 19: A second candidate signal is seen in addition to the local particle activity described above. A shorter event occurs 109 seconds before $t_{FRB}$ for a duration of 10 seconds with a logLR of 13.6. However, given the disagreement with the FRB location (Figure \ref{fig:skymaps_c}) and the number of trials, we consider this signal both insignificant and unrelated. \\

\end{document}